\documentclass{aa}
\usepackage{natbib}
\usepackage{times,graphics,latexsym}
\bibpunct{(}{)}{;}{a}{}{,}
\begin{document}

%\thesaurus{04 (08.01.1; 09.03.01; 10.01.1; 10.05.1; 10.08.01)}
\title{Chemical abundance patterns -- fingerprints of nucleosynthesis in 
the first stars}
\author{T. Karlsson
\and B. Gustafsson}
\offprints{T. Karlsson (Torgny.Karlsson@astro.uu.se)}
\institute{Department of Astronomy and Space Physics, Uppsala Astronomical 
Observatory, Box 515, SE-751 20, Uppsala, Sweden}
\date{Received 20 June 2001 / Accepted 28 August 2001}
\titlerunning{Chemical abundance patterns}
\abstract{    
The interstellar medium of low-metallicity systems undergoing star formation 
will show chemical abundance inhomogeneities due to supernova events enriching 
the medium on a local scale. If the star formation time-scale is shorter than 
the time-scale of mixing of the interstellar matter, the inhomogeneities are 
reflected in the surface abundances of low-mass stars and thereby detailed 
information on the nucleosynthesis in the first generations of supernovae is 
preserved. Characteristic patterns and substructures are therefore expected 
to be found, apart from the large scatter behaviour, in the distributions of 
stars when displayed in diagrams relating different element abundance ratios. 
These patterns emerge from specific variations with progenitor stellar mass of 
the supernova yields and it is demonstrated that the patterns are insensitive 
to the initial mass function (IMF) even though the relative density of stars 
within the patterns may vary. An analytical theory of the formation of patterns 
is presented and it is shown that from a statistical point of view the 
abundance ratios can trace the different nucleosynthesis sites even when mixing 
of the interstellar medium occurs. Using these results, it should be possible 
to empirically determine supernova yields from the information on relative 
abundance ratios of a large, homogeneous sample of extremely metal-poor 
Galactic halo stars.      
}
\maketitle

\section{Introduction}
\label{intro}
\noindent
During the last decade, the methods for abundance determinations for faint
Pop. II stars have reached a stage when accurate abundance ratios can be
obtained for great samples of stars with overall metallicities below $1/100$,
and even $1/1000$, of the solar. This has made it possible to explore not
only abundance trends, e.g. the variation of oxygen or magnesium abundances
with decreasing iron abundance, but also the intrinsic scatter in abundance 
ratios such as Eu/Fe, at a given Fe/H as a function of Fe/H. 

\par    

An early example of a detailed discussion of the dispersion in abundance
ratios was that of Edvardsson et al. (1993)\nocite{eetal93} in their study of 
the chemical evolution of the Galactic disk. These authors found a significant 
scatter in Fe/H for Disk stars of a given age and a given galactocentric mean
distance; however, they did not find any tendency for e.g. Mg/Fe or other
$\alpha$-element/iron abundance ratios to scatter at a given overall 
metallicity, except for the most metal-poor stars. A tendency for these 
latter stars could be interpreted as the result of a greater star-formation 
rate (SFR) in the inner Galaxy. In their study of Pop. II dwarfs 
Nissen at al. (1994)\nocite{netal94} found that the scatter in abundance ratios 
Mg/Fe, Ca/Fe or Ti/Fe for their sample of Galactic halo stars with 
$-3.2 \le$[Fe/H] $\le -1.8$ was less than $0.06$ dex. They also found an upper 
limit in the scatter in O/Fe of $0.15$ dex. Since the yields of these different 
elements are different from supernovae (SNe) of different masses, the authors 
could conclude from the small abundance scatter that the elements observed in 
the stars must be the results of at least about 20 SN explosions, otherwise 
statistical fluctuations in abundances should have been present. 

\par

The fact that a scatter in abundance ratios for the most metal-poor stars
should result from a small number of SNe, with different masses and therefore
different yields of heavy elements, was also pointed out by Audouze \& Silk
(1995)\nocite{as95}. McWilliam et al. (1995)\nocite{mcwilliam95} explained the 
large scatter in s-process element abundances like Ba/Fe and Sr/Fe for 
[Fe/H]$\le-2$ in similar terms (see also McWilliam et al. 1996;\nocite{metal96} 
McWilliam 1998\nocite{mcwilliam98}; Ryan et al. 1996\nocite{ryan96}).

\par

The abundance scatter for Halo stars has also been modelled. In a stochastic
Halo formation model Argast et al. (2000)\nocite{argast00} studied the scatter 
in relative abundances as resulting from the small number of SNe contributing 
to the abundances for the most metal-poor stars; these authors found that a 
great scatter should be expected in several ratios for stars with [Fe/H]$<-3$, 
representing early evolutionary phases when the interstellar medium (ISM) of 
the Halo was unmixed and dominated by local inhomogeneities. In the
range where [Fe/H] increases from $-3$ to $-2$ there is a gradually 
diminishing scatter due to the contribution from an increasing number of SNe
and more mixing in the ISM, while for still higher metallicities the Halo ISM
is well mixed, and homogeneous abundance ratios result. Recent studies, 
empirical as well as theoretical, of chemical inhomogeneities in the Galactic 
halo have also included work on r-process elements (notably Eu, as well as Ba,
see Ishimaru \& Wanajo 1999\nocite{ishimaru99}; Raiteri et al. 
1999\nocite{raiteri99}; Travaglio et al. 2001\nocite{travaglio00} and
references given there) and Be and B 
(e.g. Suzuki et al. 1999\nocite{suzuki99}). In particular, Tsujimoto et al. 
(2000\nocite{tsy00} and references therein) have developed a stochastic 
chemical evolution model to study the effects of inhomogeneous r- and 
s-process element enrichment in the early Galaxy. Some of their relative 
abundance diagrams show features resembling those to be discussed here.

\par

There are today three identified metal-poor Halo stars with dramatic 
r-process signatures where CS 22892-052 is the most famous one 
(Sneden et al. 1996\nocite{sneden96}). The identification of such extreme 
outliers gives important clues to the understanding of the nucleosynthesis in 
these early SN explosions. However, it is not possible to determine, from the outliers alone, which type of SN (i.e. what progenitor mass) is able to produce 
such an abundance signature, nor to estimate the relative significance of 
different SNe with other progenitor masses. In this paper, we shall present an 
alternative approach to unveil the statistical properties of SNe, taking into 
account a whole population of extremely metal-poor stars. The effects from a
small number of SNe affecting the chemical composition of these stars are 
explored.
We shall demonstrate the probable presence of fine structure patterns in the 
diagrams where abundance ratios are plotted relative to each other, reflecting 
the number of SNe contributing and the sometimes strong mass dependence of the 
yields. In fact, we shall argue that these patterns, if observed, could be used 
for empirically exploring such properties further. Some preliminary results of 
our work were published in Karlsson \& Gustafsson (2000)\nocite{kg00}. 

\par

After some general comments and definitions in Sect. \ref{sec2}, we shall 
present our simulations in Sect. \ref{sim}. The origin of the patterns is 
further analysed in Sect. \ref{math}, wherein an analytical theory of the 
distributions of stars in different abundance planes is developed. 
Observational implications are discussed in Sect. \ref{discussion} and the 
conclusions are presented in Sect. \ref{concl}. In the Appendix, we derive some 
general expressions describing the statistics of random variables.

\section{Chemical enrichment in metal-poor systems}
\label{sec2}
\noindent
Local chemical inhomogeneities in the interstellar medium, caused by the first
generations of SNe in the Galaxy, may or may not be preserved in subsequent 
generations of stars, depending on how efficient the mixing of the ISM is 
relative to the rate of star formation. The global mixing can be 
defined in terms of a mixing efficiency parameter, 
$\epsilon_{\mathrm{mix}}$, which describes how fast the mixing 
occurs as measured in, say, solar masses per million years. Hence,  
$\tau_{\mathrm{mix}} \propto M_{\mathrm{system}}/\epsilon_{\mathrm{mix}}$, 
where $M_{\mathrm{system}}$ is the total mass of the system. The mixing 
efficiency parameter can also be used to define a characteristic 
mixing mass, such that $M_{\mathrm{mix}}=\epsilon_{\mathrm{mix}} 
\tau_{\mathrm{SF}}$, where $\tau_{\mathrm{SF}} \propto \mathrm{SFR}^{-1}$ is
the star formation time-scale.

\par

Thus, if the mixing is efficient enough or/and the star-formation rate is low, 
the time-scale of global mixing is much shorter than the star
formation time-scale, or equivalently, $M_{\mathrm{mix}}$ is comparable to 
$M_{\mathrm{system}}$. In this case the inhomogeneities are wiped out before subsequent generations of stars are formed and the system will be considered 
well mixed at any instant of time. The stars will reflect the cumulative 
build-up of the elements and the whole stellar population has, in this case, a 
common chemical history meaning that a time-axis can be defined. This, in turn, 
implies the existence of an age-metallicity relation. On the other hand, if 
$\tau_{\mathrm{mix}} \gg \tau_{\mathrm{SF}}$, i.e., if $M_{\mathrm{mix}}\ll 
M_{\mathrm{system}}$ the stars would trace the local fluctuations caused by 
individual enrichment events and the chemical inhomogeneous state of the 
system would be resolved and preserved by low-mass (yet unevolved) stars. Such 
a system can not be considered evolving with time in a unique way since the 
stars in the system describe different chemical histories. Hence, a single 
age-metallicity relation does not exist.

\begin{figure}
 \resizebox{\hsize}{!}{\includegraphics{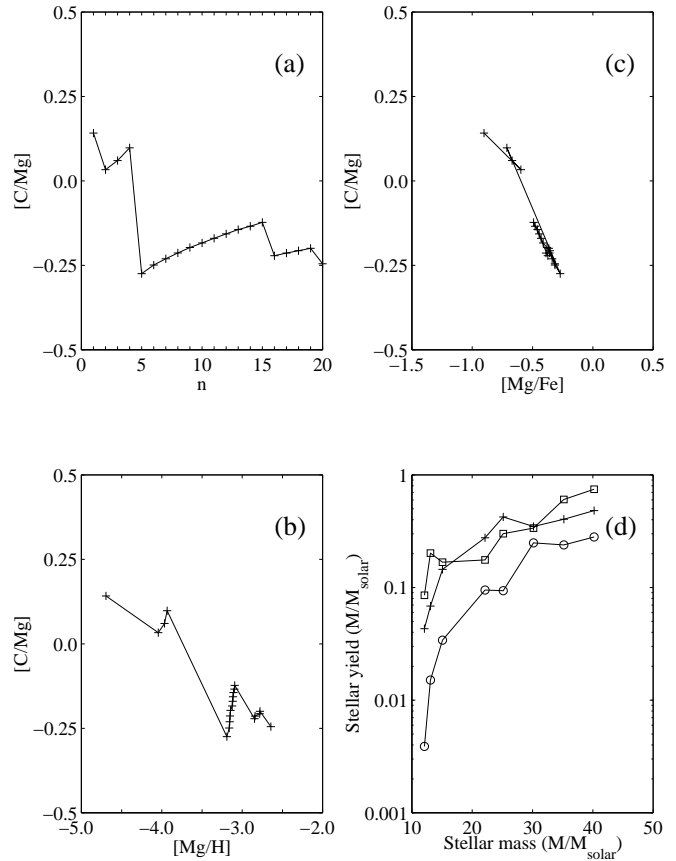}}
 \caption{\textbf{a)} A sequential build-up of the C/Mg ratio.
  The ratio changes when the contribution from another SN is added. The
  number of SNe is denoted by $n$. \textbf{b)} The chemical track in the 
  [Mg/H]--[C/Mg] plane. The amount of magnesium is mixed with 
  $5\times 10^{5}~\mathcal{M_{\odot}}$ of hydrogen in calculating the Mg/H 
  ratio. \textbf{c)} The chemical track in the [Mg/Fe]--[C/Mg] plane. 
  \textbf{d)} The variation of the stellar yields with SN mass taken from the 
  zero-metallicity models of Woosley \& Weaver 
  (1995). The pluses ($+$) denote the carbon yield, the open circles 
  ($\circ$) denote the magnesium yield and the squares ($\Box$) denote the 
  iron yield. Note the downturn in the carbon yield at 
  $\sim25~\mathcal{M_{\odot}}$. At the same mass, the Mg-yield 
  increases. This is reflected by the jumps in the chemical tracks}
 \label{basic}
\end{figure}

\par

Let us discuss in more detail how the chemical enrichment of the ISM can be 
reflected in the stellar population. Suppose that we have a system which 
initially consists of primordial gas (i.e. $Z_{\mathrm{init}}=0$). Assume that
the instantaneous recycling approximation is valid and that the whole system 
is well mixed at all times, i.e. $\tau_{\mathrm{mix}} \ll \tau_{\mathrm{SF}}$. 
Any massive star that explodes immediately pollutes the whole system and the 
chemical composition of the ISM is changed accordingly. Let us call a chain of 
such enrichment events a chemical series. Note that the progenitor masses
of the SNe do not follow a decreasing sequence, they are randomly distributed.
A finite part of a series, say, up to $i$ SNe, shall be referred to as a 
chemical sequence which describes the chemical state of the system at the 
time when $i$ SNe have exploded. A star formed out of this gas 
is a realization of the chemical sequence. Now, if low-mass stars are 
continuously formed over time the complete enrichment history of the ISM is 
mapped. The chemical evolution of the system is then followed by 
displaying the stars in different abundance diagrams. Such time-lines 
shall be called chemical tracks and are realizations 
of the chemical series up to a certain number, $n$, of SNe. Since the ISM is 
homogeneous there exists only one chemical series in the system. This is 
illustrated in Fig. \ref{basic} by the sequential build-up of the C/Mg ratio. 
Two chemical tracks related to the C/Mg ratio are also shown 
(Fig. \ref{basic}b, \ref{basic}c). The distinct jumps in Figs. \ref{basic}a 
and \ref{basic}b reflect the enrichment of the ISM by a massive SN. The yields 
as functions of progenitor mass are shown in Fig. \ref{basic}d. 
 
\par

Systems like the one discussed above show unique chemical tracks for each
pair of elements (or element ratios). This is clear since the chemical state
of such a system at any time is described by a single chemical series. Suppose 
now that $M_{\mathrm{mix}} \ll M_{\mathrm{system}}$, i.e. our system 
consists of a number of separate subsystems, or star-forming regions. As above, 
the instantaneous recycling approximation is valid for each region but the 
different subsystems evolve differently, and their chemical states are 
described by different chemical series. A population of stars, randomly 
picked from different star-forming regions would, then, describe the 
chemical evolution of several systems, leading to a large star-to-star scatter 
in the observed abundances. However, due to the many different enrichment 
histories of these stars, substantially more information on the elemental 
production sites (described as stellar yields) is preserved and we shall 
see that the distribution of stars in the abundance diagrams show structures 
and patterns created by specific variations in the SN yields with progenitor 
mass which can provide detailed clues to the production of elements in the 
early Galaxy.

\section{Simulations of the chemical patterns}
\label{sim} 
The existence of a large scatter in abundance ratios for the Galactic halo 
stars with [Fe/H]$<-2.5$ suggests the second scenario 
(i.e. $\tau_{\mathrm{mix}} \gg \tau_{\mathrm{SF}}$) discussed in 
Sect. \ref{sec2}. Below, we shall simulate the chemical enrichment of such a 
metal-poor system. In all our simulations particular data on SN yields will be 
applied. These data are rather uncertain, not the least as regards the 
variations of the yields with stellar mass, which also makes our actual 
predicted abundance patterns uncertain. Our point here is, however, not to make 
exact predictions but to illustrate the general effects of mass-dependent 
yields. 
 
\subsection{General assumptions}
\label{assumptions}
\noindent
The gaseous medium of the system (i.e. the Galactic halo) is assumed to be 
initially primordial, i.e., $Z_{\mathrm{init}}=0$. The characteristic mixing 
mass is assumed to be much smaller than the total mass of the system, such that 
$M_{\mathrm{mix}}\ll M_{\mathrm{system}}$. This is realized by assuming 
$M_{\mathrm{mix}}$ to be of the same order of magnitude as the localized 
star-forming regions of, say, $\sim10^6~\mathcal{M_{\odot}}$ each. From the 
number of Halo horizontal branch stars, 
Binney \& Merrifield (1998)\nocite{bm98} estimate the present day mass of the stellar component of the Halo to $\sim 10^{8}~\mathcal{M_{\odot}}$, which implies that the total mass of the Halo was, at least, on the order of 
$M_{\mathrm{system}}=10^{9}~\mathcal{M_{\odot}}$ in the early epochs. Thus, 
$M_{\mathrm{mix}}/M_{\mathrm{system}} \sim 10^{-3}$. 

\par

We shall distinguish between abundances measured relative to hydrogen, and 
abundance ratios between different, more heavy elements. To trace individual 
SNe or groups of SNe using a ratio of some element $A$ relative to hydrogen (an
$A$/H ratio), requires knowledge about the mixing of the star-forming region after the SNe ejection. Ratios between heavier elements, both produced in 
a SN, are less sensitive to the mixing. This is due to the fact that an 
$A$/H ratio, in the first-order expectation, essentially depends on the mean 
distance from the contributing SNe while ratios between different, heavier elements do not. In our models we first schematically assume that the newly 
synthesized elements mix with a constant hydrogen mass, i.e. all star-forming 
regions are supposed to have equal mass. Furthermore, the mixing with the cloud 
material is assumed to be instantaneous and complete. Subsequently, low-mass 
star formation occurs within each cloud. This implies that the time-scale of 
mixing within the region is shorter than the formation time-scale for 
individual stars which in turn is shorter than the epoch of the star formation 
activity in the region and the life-time of the region itself.   

\par

Recurrent SN explosions may disrupt the regions before local mixing and 
subsequent star-formation occurs. If this happens inter-cloud mixing has to be 
taken into account. This will naturally affect abundances relative to hydrogen 
in the ISM. However, abundance ratios between more heavy elements are, at 
least statistically, almost unaffected by such a mixing. 

\par 

Subsequently, we shall discuss two types of abundance-ratio diagrams (in the 
general case denoted "$A/$ diagrams" below). Some diagrams will relate the 
abundances of two heavier elements to hydrogen, such as [C/Mg] vs. [Mg/H]; 
these are  denoted "$A/$H diagrams". These diagrams will be affected by the 
not very well-known degree of mixing of SN remnants with other, less processed, 
material in the different star-forming regions. Therefore, we shall primarily 
discuss diagrams displaying the relative abundances between three heavier 
elements, e.g. [C/Mg] vs. [Mg/Fe], diagrams that are much less sensitive to 
the mixing uncertainties. This type of diagrams will be denoted 
"$A/A$ diagrams" below.

\par

The ejected material from each individual SN is assumed to be well mixed 
prior to subsequent star formation, such that any stars formed out of this 
material will all have the same abundances.
Both observations (see Travaglio et al. 1999\nocite{travaglio99}) and 
hydrodynamical simulations (e.g. Kifonidis et al. 2000\nocite{kifonidis00}) of 
core collapse supernovae show evidence for substantial mixing of elements 
shortly after the core bounce (however, see also Hughes et al. 
2000\nocite{hughes00}; Douvion et al. 1999\nocite{douvion99} where they 
discuss heterogeneous mixing in the Cas A supernova remnant). The simulations 
also indicate that the interaction between the processed material and the outer 
layer of hydrogen (i.e in SNe type II) causes a complete homogenization of the 
ejected material. This process does not work for stars without a massive 
hydrogen envelope (i.e. SNe Ib/Ic) for which the mixing might be less 
pronounced (Kifonidis, private communication).          

\par

In our investigation we assume that the stellar yields are one-dimensional 
functions of progenitor mass of the SN. For convenience, we 
also assume that stars less massive than $10~\mathcal{M_{\odot}}$ do not 
produce heavy elements within the time-scales considered here and that the
prompt enrichment of the primordial ISM by a population of very massive stars 
($>100~\mathcal{M_{\odot}}$) suggested by, e.g., Wasserburg \& Qian 
(2000)\nocite{wq00} did not occur. Recent yield calculations by 
Umeda et al. (2000)\nocite{umeda00} show a moderate metallicity dependence in the lowest metallicity regime for the secondary elements, notably $^{14}$N, 
$^{23}$Na and $^{27}$Al. However, primary elements such as $^{12}$C, 
$^{16}$O and $^{24}$Mg are almost independent on metallicity. A high 
dependence enters first after the amount of metals initially present is 
sufficient to cause extensive mass-loss through radiation-driven winds 
(Maeder 1992\nocite{maeder92}; Portinari et al. 1998\nocite{petal98}). The 
chemical yields may also be altered by stellar rotation 
(Heger \& Langer 2000\nocite{hl00}). It is straight forward to incorporate 
rotational dependent yields in the simulations, as well as in the analytical theory presented below (Sect. \ref{anal_app}). The situation is different for 
metallicity dependent yields since the metal content in every star is coupled 
to the history of chemical enrichment. We shall neither consider metallicity- 
nor rotational dependent yields in the present study.

\par

For the general discussion we shall assume that the amount of any heavy element 
ejected by a SN of a given progenitor mass is constant from one stellar 
generation to the next. The ejecta of an element $A$, $e_A(m,Z)$, can be 
written as 

\begin{equation}
  e_A(m,Z)=(m-m_r) \times Z_A^{\mathrm{init}}(t-\tau_{m})+p_A(m),
  \label{ejecta}
\end{equation}

\noindent
where $m$ is the progenitor mass of the star, $m_r$ is the remnant mass, 
$Z_A^{\mathrm{init}}(t-\tau_{m})$ is the initial mass fraction of the element 
at the time of formation of the star, $\tau_{m}$ is the life-time of stars
with mass $m$ and $p_A(m)$ is the stellar yield, the mass of element $A$ 
produced or destroyed in the star. Hence, the total ejected mass is a sum of 
the initially present amount of $A$ and the newly synthesized amount. For low 
metallicities the first term on the RHS in Eq. (\ref{ejecta}) becomes 
negligible and, thus, the ejected matter is completely dominated by the 
stellar yield. 
 
\par

We assume the sampled Halo stars to be statistically independent. Thus, we 
assume that all chemical sequences (i.e. stars) are picked randomly from an
infinite number of chemical series. This assumption means physically that the 
stars that originate from star formation regions where, say, three SNe have 
exploded sample different such regions, and that stars coming from regions 
where four SNe have exploded sample still other and different regions. This 
assumption is further discussed in Sect. \ref{stat_indep} and there found to be 
reasonable.

\subsection{Model I}
\noindent
To simulate the distribution of abundances in Halo stars we adopt the 
simple picture that star formation and heavy element enrichment of the ISM are 
confined within star-forming regions of about a Jeans mass each, i.e. a 
hydrogen mass of $M_{\mathrm{H}}=5\times10^{5}~\mathcal{M_{\odot}}$. This 
particular value is chosen {\it ad hoc}. Another hydrogen mass would simply 
introduce a shift in the abundances relative to hydrogen. We place a number of 
high-mass stars with randomly distributed masses (according to the Salpeter IMF 
if nothing else is stated) between $10$ and $100~\mathcal{M_{\odot}}$ in the 
hydrogen clouds and let them explode as core collapse SNe. The total number of 
SNe exploding in each cloud vary in a range from one to a large number (i.e. 
up to $\sim 20$). The SNe produce heavy elements according to theoretically 
calculated stellar yields. We shall use the yields by 
Woosley \& Weaver (1995)\nocite{ww95} 
calculated from their metal-free models (Z models) and by Nomoto et al. 
(1997)\nocite{netal97}. Hereafter, these works will be referred to as WW95 and 
Netal97, respectively. All yields are extrapolated by a constant to the ends of 
the mass interval. The yields by WW95 are modified according to the possible 
decay channels of the unstable isotopes. The high-mass stars explode directly 
and their ejecta are instantaneously mixed with the cloud material. Hence, 
these regions are chemically different according to the number and masses of 
SNe that have formed and enriched the regions. Subsequently, low-mass stars are 
formed out of the enriched gas. Effectively, this means that we sum up the 
heavy element contribution from every SN within each cloud. The abundance of 
any element $A$ relative to hydrogen (by number) is then calculated and taken 
as the surface abundance of a low-mass star formed in the cloud. 

\par

We assume the total number of low-mass stars formed in each cloud to be 
constant. This implies that there is an equal probability to pick a star 
tracing one chemical series as it is to pick a star tracing another series, 
and since the number of star-forming regions is assumed to be large 
(i.e. the stars are statistically independent) we are allowed to select the 
stars from different regions. The total number of Halo stars ($N$) in a 
simulated sample is then governed by  

\begin{equation}
  N=\sum\limits_{i=1}^n W_i,
  \label{N_tot}
\end{equation}

\noindent
where $W_i$ is the number of clouds in which $i$ SNe explode and $n$ is 
the maximum number of SNe that is allowed to explode and enrich a 
single cloud. In a typical simulation we assume $W_i$ to be constant, e.g. 
$W_i=W=25$ for all $i$, thus the total number of stars is $N=n \times W$. By 
displaying various abundance ratios for low-mass stars in different $A/$ 
diagrams we are then able to follow the early chemical enrichment phase of an 
initially metal-free system.

\begin{figure}
 \resizebox{\hsize}{!}{\includegraphics{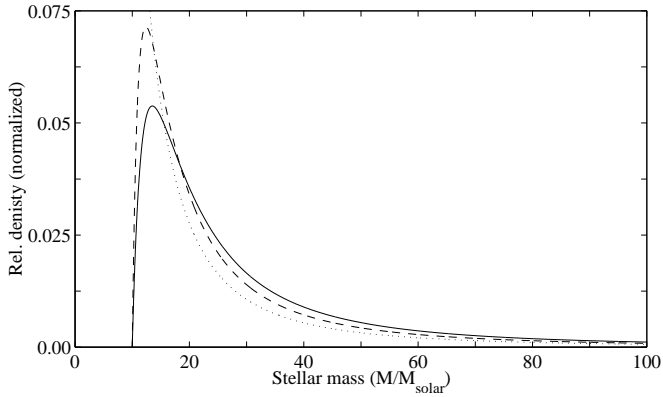}}
 \caption{Different mass distribution functions as described by 
  Eq. (\ref{esfmdf}). The full line denotes the relative number density of 
  exploded stars at $t=28.7$ Myrs (corresponding to the life-time of a 
  $10~\mathcal{M_{\odot}}$ star) for a constant SFR. The dashed line denotes 
  the relative number density for an exponential SFR with 
  $t_{\mathrm{SF}}=10$ Myrs. The dotted line is the Salpeter IMF}
 \label{mf}
\end{figure}

\subsection{Model II}
\noindent  
We have alternatively modified our simple chemical enrichment model by 
introducing continuous star formation in the regions. This mainly 
affects two parameters, the mass distribution function (which is, in Model I, 
equal to the IMF) and the lower mass limit of this distribution. 

\par

In Model I, the high-mass stars are formed in an initial burst. By allowing 
the stars (both high- and low-mass stars) to form continuously over a certain 
period of time we can account for a mass distribution function that changes 
with time as stars with different masses have different life-times. We shall 
adopt an exponential star formation rate ($\psi$) such that 

\begin{equation}
\psi(t)=\psi_{0}e^{-t/t_{\mathrm{SF}}},
\end{equation}

\noindent
where $t_{\mathrm{SF}}$ is a characteristic time of the star formation. This
time is set to the same value in all clouds. Furthermore, suppose that the IMF 
is a power law of the form $\phi(m)=\phi_{0}m^{-\alpha}$ and introduce 
$\tau_{m}$, the life-time of a star with mass $m$. The expression for the 
time-dependent mass distribution function (see Eq. (7.4) in 
Pagel 1997\nocite{pagel}) describes the distribution of stars over $m$ at a certain $t$. However, we are more interested in the stars that have enriched 
the cloud at time $t$. The corresponding mass distribution of exploded stars, 
(see Fig. \ref{mf}) is governed by 

\begin{equation}
  f(m,t) = \left\{ \begin{array}{ll}
     0, & \tau_{m}>t \\
     f_{0} m^{-\alpha}
     (1-e^{-(t-\tau_{m})/t_{\mathrm{SF}}}), & \tau_{m}\le t,
  \end{array} \right.
  \label{esfmdf}
\end{equation} 

\noindent
where $f_{0}=\phi_{0}\psi_{0}t_{\mathrm{SF}}$. The stellar life-times are adopted from the lowest metallicity models of Portinari et al. 
(1998)\nocite{petal98}. Note that in Model I, the function $f(m,t)$ was 
equal to the IMF since all the high-mass stars were allowed to explode before the formation of the low-mass stars.

\par

The star formation period is assumed to be $\sim 30~\mathrm{Myrs}$ in each 
cloud. This is only slightly longer than the estimate by 
Shull \& Saken (1995)\nocite{ss95}, for OB associations. A lower life-time of 
the clouds would hinder the formation of stars that could be enriched in 
elements produced by the least massive SNe, as our models do not take into 
account global mixing and a second generation of star-forming regions. 

\par

Except for the continuous star formation rate Model II is based on the same 
assumptions as Model I. We adopted a constant star-formation rate in each 
cloud, which corresponds to a long characteristic time, $t_{\mathrm{SF}}$ 
(see Fig. \ref{mf}), leading to a mass distribution of SN progenitors 
that is significantly different from the IMF. The slope of the IMF was set to 
$\alpha=2.35$. Furthermore, a read-off time ($t_{\mathrm{read-off}}$),
distributed according to the SFR, was generated in each cloud which determined 
the actual number of polluting SNe, $i$, via an integration of the SFR (up to 
$t_{\mathrm{read-off}}$) normalized to the total number of high-mass stars in 
the cloud. The total number of high-mass stars formed in each cloud was 
randomly generated according to a Gaussian distribution centred at $n=10$ and 
with $\sigma=20$ (for $n<0$ the probability is zero). Thus, the number of 
stars that have  been enriched by $i$ SNe ($W_i$) is not {\it a priori} known 
but is determined after the simulation stops. As mentioned above, this read-off 
time also sets the lower cut-off of the distribution function of the exploded 
high-mass stars as no star with a longer life-time than $t_{\mathrm{read-off}}$
has been able to enrich the cloud.

\subsection{Results}
\label{results}
\noindent
We should emphasize that a realistic modelling of the early chemical enrichment
requires a more physical treatment of the mixing than adopted here. The 
abundances relative to hydrogen are particularly sensitive. The situation is 
quite different for abundance ratios, however. They are independent of 
variations in the mixing mass. They are also insensitive to global, inter-cloud 
mixing, infall and subsequent generations of star-forming regions (see the 
discussion in Sect. \ref{ratio}). Therefore, we shall, based on this 
distinction, separately discuss the two corresponding types of $A/$ diagrams. 

\begin{figure}
 \resizebox{\hsize}{!}{\includegraphics{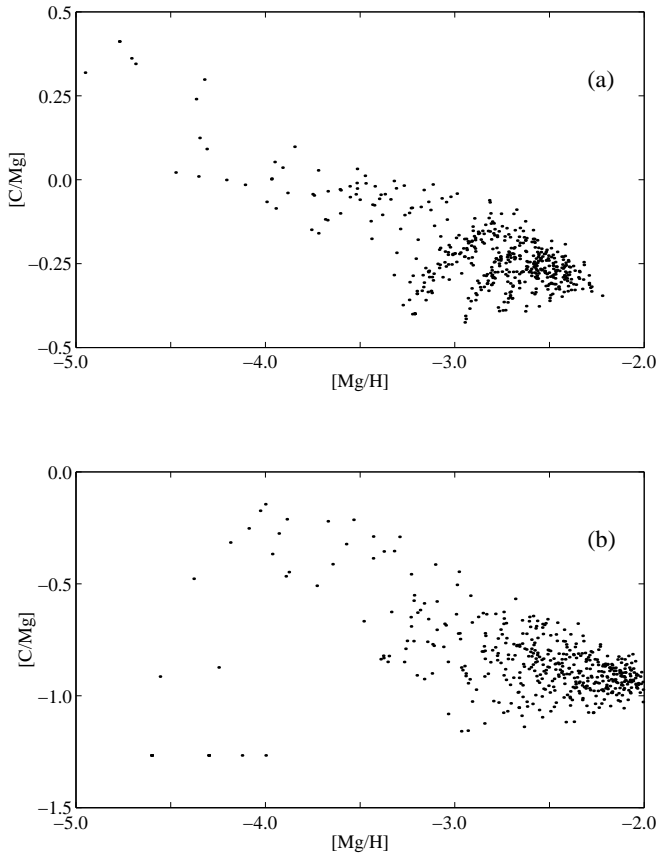}}
 \caption{\textbf{a)} A scatter plot of simulated stars in the [Mg/H]--[C/Mg] 
  plane (Model I). The diagram shows $500$ low-mass stars enriched by $1-20$ 
  SNe ($W_i=25$). The yields are taken from WW95 and the masses of the SNe, 
  ranging from $10~\mathcal{M}_{\odot}$ to $100~\mathcal{M}_{\odot}$, are 
  distributed according to a Salpeter IMF. \textbf{b)} A scatter plot of stars 
  in the same plane as above, using yields from Netal97. It is seen that the 
  existence of clear patterns in these $A/$H diagrams are dependent on the 
  details of the predicted SN yields}
 \label{CMg_fig}
\end{figure}

\begin{figure}
 \resizebox{\hsize}{!}{\includegraphics{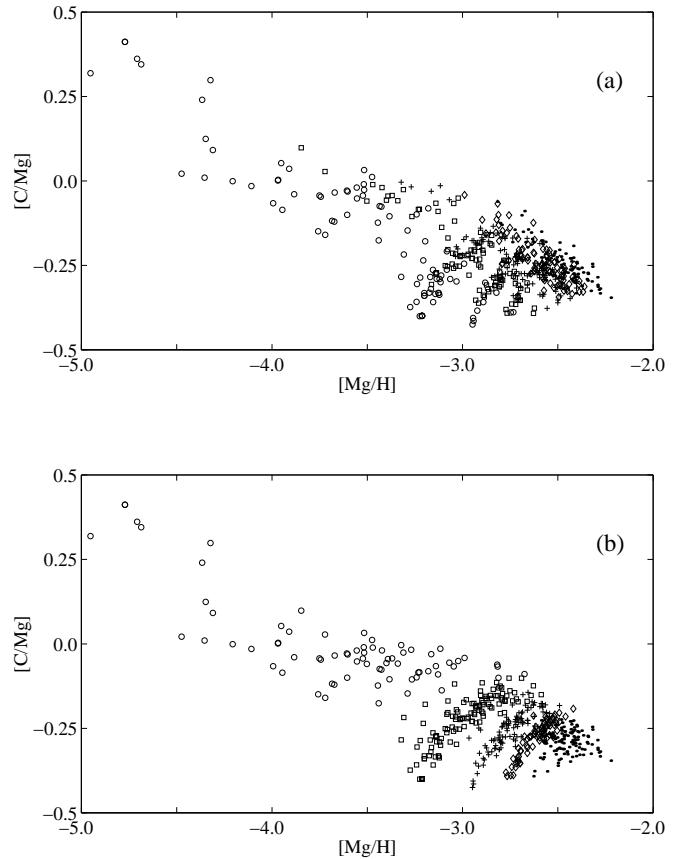}}
 \caption{\textbf{a)} Same as Fig. \ref{CMg_fig}a except that the stars are 
  coded. Circles ($\circ$) denote stars enriched by $1-4$ SNe, squares ($\Box$)
  are enriched by $5-8$ SNe, pluses ($+$) by $9-13$ SNe, diamonds ($\Diamond$) 
  by $14-17$ and dots ($\cdot$) are enriched by more than 17 SNe. \textbf{b)} 
  As above but with a different coding. Circles ($\circ$) denote stars 
  which have not been enriched by any SN with a mass above 
  $27.5~\mathcal{M}_{\odot}$, squares ($\Box$) are enriched by one SN above 
  $27.5~\mathcal{M}_{\odot}$, pluses ($+$) have been enriched by two such SNe, 
  diamonds ($\Diamond$) by three and dots ($\cdot$) by four or more SNe above
  $27.5~\mathcal{M}_{\odot}$. The patterns displayed reflect the number of 
  high mass SNe that have enriched the ISM before the sample star was formed}
 \label{CMg_coded_fig}
\end{figure}

\subsubsection{$A/$H diagrams}
\noindent
Let us first discuss some properties of the $A/$H diagrams before we turn 
to the $A/A$ diagrams. We see from Fig. \ref{CMg_fig} that the appearance
of the stars in the diagrams depends naturally on the produced amount of 
carbon and magnesium in the massive stars. However, the shape of the 
stellar yield functions are responsible for possible trends and/or the 
groupings of stars into different substructures and patterns. These structures 
appear as a result of the various enrichment histories of the low-mass stars 
(cf. Fig. \ref{basic} and the discussion in Sect. \ref{sec2}). Even though we 
generally do not expect patterns to appear in $A/$H diagrams displaying real 
observations the overall distribution of stars, such as large-scale trends, 
can still hold important information. If patterns would really be observed in 
$A/$H diagrams this would put strong constraints on the star formation 
and mixing processes in the early Galaxy. There are three effects that are 
directly observed in the diagrams in Fig. \ref{CMg_fig} and Fig. \ref{CMg_coded_fig}.

\par

Firstly, the star-to-star scatter seems to decrease with increasing 
metallicity (represented by [Mg/H]). This is best seen in Fig. \ref{CMg_fig}b 
and is due to the fact that stars enriched by a single SN have the 
lowest metallicities and the largest variations in the C/Mg 
ratio. When more and more SNe contribute to the metal content in the
low-mass stars the metallicity increases and the variation in the ratio is
averaged out. Note also that stars with a specific metallicity may have been 
enriched by quite a different number of SNe. For example, at [Mg/H]$\sim -3$ 
the stars with lowest C/Mg ratio have been enriched by perhaps a couple of SNe 
while the ones with the highest ratio have been enriched by up to $17$ SNe 
(see Fig. \ref{CMg_coded_fig}a).   

\par

Secondly, in Fig. \ref{CMg_fig}a we see another effect. Instead 
of a decreasing scatter with metallicity the scatter is 
asymmetric, mimicking a trend. The C/Mg ratio seems to decrease with increasing 
Mg/H. This is not a normal evolutionary effect caused by time or metallicity 
(such as the decrease of [Mg/Fe] with [Fe/H] for [Fe/H]$>-1$ induced by the 
onset of thermonuclear supernovae (SNe type Ia) or the increase of [C/O] with 
[O/H] which could be explained by a metallicity dependent carbon yield as 
proposed by e.g. Gustafsson et al. 1999; \nocite{gustafsson99} Henry et al. 
2000\nocite{henry00}) but rather a SN mass (i.e. number) effect. This is 
accomplished by SNe producing a high C/Mg ratio at the same time 
produce a small amount of magnesium while SNe producing a low C/Mg ratio also 
produce much Mg. So, for an extremely metal-poor system an observed trend like 
this one does not necessarily imply chemical {\it evolution} in the normal 
sense (see also Tsujimoto \& Shigeyama 1998\nocite{ts98}).           

\par

Thirdly, the stars in Fig. \ref{CMg_fig}a tend to group together in
substructures. It is understood from Fig. \ref{CMg_coded_fig}b that these
patterns are caused by a specific variation with progenitor mass in the carbon 
yield. Roughly, one can say that these patterns are formed by a pronounced 
decrease in the yield around $27.5~\mathcal{M}_{\odot}$. We shall discuss these
issues in more detail in Sect. \ref{math}. As we have mentioned, the 
substructures in Fig. \ref{CMg_fig}a are sensitive to variations in the mixing 
mass. As long as all star-forming regions have equal mass these substructures 
survive. This is unlikely, however. Nakasato \& Shigeyama 
(2000)\nocite{nakasato00} discuss metal enrichment of the primordial ISM by 
individual SNe and find that the Mg/H ratio in various filaments may differ
by $\sim 1.0$ dex, implying a large abundance scatter in the second generation 
of stars. Thus, a smearing 
effect most probably occurs in the horizontal direction in the diagrams of 
Figs. \ref{CMg_fig} and \ref{CMg_coded_fig}. The fact that we see patterns in 
the $A/$H diagrams is because we have not included this type of mixing in 
the models. 

\begin{figure*}[t]
 \resizebox{\hsize}{!}{\includegraphics{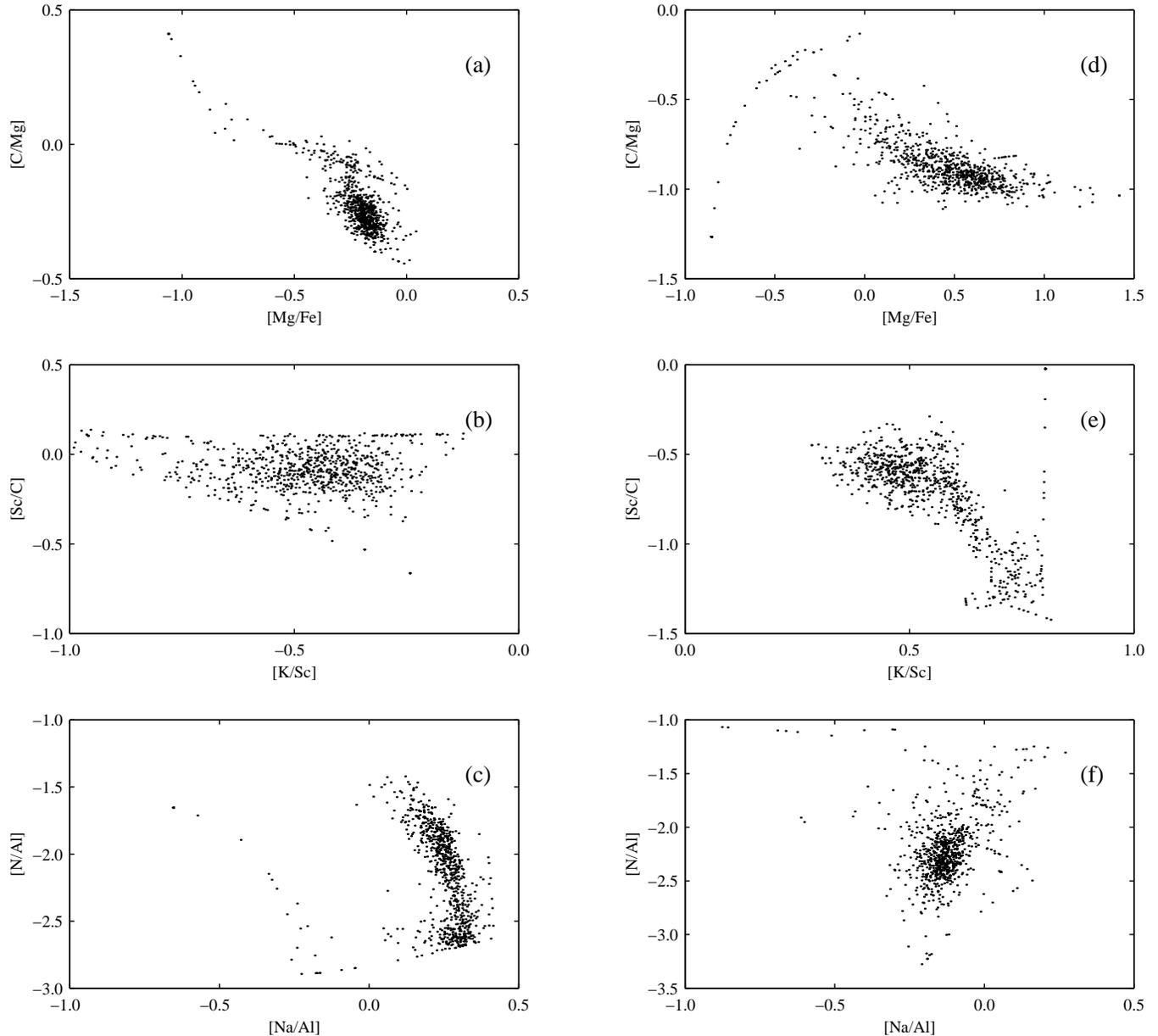}}
 \caption{Scatter plots of stars as they may appear in different $A/A$ 
  diagrams. \textbf{a}-\textbf{c)} Simulated stars (Model I) 
  enriched by 1 -- 20 SNe using yields from WW95. 
  \textbf{d}-\textbf{f)} The corresponding $A/A$ diagrams produced with yields 
  taken from Netal97. Note that for a particular abundance plane the 
  difference (apart from the loci) between the distributions is large which
  arises from differences in the yields}
 \label{corr_diagrams}
\end{figure*}

\subsubsection{$A/A$ diagrams}
\noindent
Using Model I, it is possible to generate pure abundance ratios which can be 
displayed in $A/A$ diagrams such as those in Fig. \ref{corr_diagrams}.
The patterns that are formed in these kinds of diagrams, opposed to the ones
in the $A/$H diagrams, are insensitive to intrinsic uncertainties such as 
mixing. They also show larger variations in their shapes. For observational 
uncertainties of $\simeq 0.1$ dex in the Mg/H and C/Mg ratios the two scatter 
plots in Fig. \ref{CMg_fig}a and b would appear quite similar apart from the 
different loci. However, for the same uncertainties in the C/Mg and Mg/Fe 
ratios the difference between the chemical patterns in Figs. 
\ref{corr_diagrams}a and \ref{corr_diagrams}d will survive. It is obvious 
that the stellar yields play a crucial role in the formation of these 
patterns. The other pairs in Fig. \ref{corr_diagrams} show even larger 
differences due to larger disagreements in the two sets of yields (i.e. WW95 
and Netal97). 

\par

\begin{figure*}[t]
 \resizebox{\hsize}{!}{\includegraphics{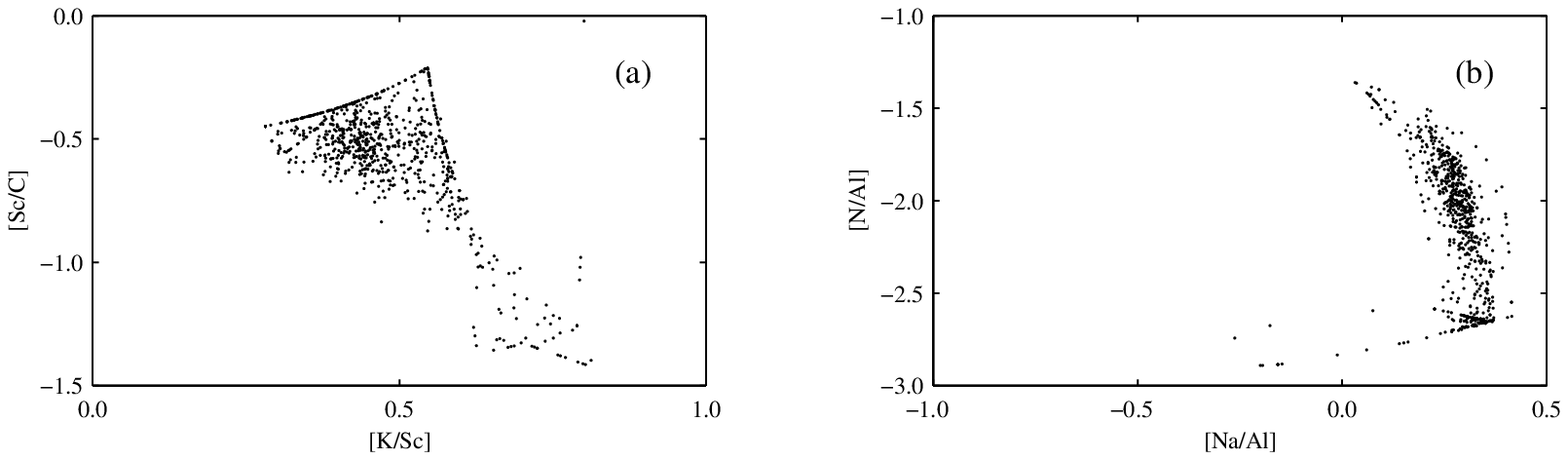}}
 \caption{\textbf{a)} The distribution of stars in the [K/Sc]--[Sc/C] plane
  (cf. Fig. \ref{corr_diagrams}e) as predicted by Model II. Due to the 
  evolutionary effect the highly enriched stars do not converge to a point in 
  the diagram even though they have been enriched by many SNe. This 
  effect is not easily detected, however. \textbf{b)} Same as Fig. 
  \ref{corr_diagrams}c as predicted by Model II}
 \label{model2_figs}
\end{figure*}

Model I and Model II generate $A/A$ diagrams showing strong similarities.
By comparing Fig. \ref{model2_figs}a and Fig. \ref{corr_diagrams}e we
see that the stars are arranged in the same manner, apart from minor 
differences in the distribution within the formed pattern. This is a result of
the discrete enrichment as the abundance ratios are not considered to be 
weighted with an IMF or some other mass distribution function. A set of yields
generates a unique pattern while the mass frequency of SNe determines
how the pattern will be populated by low-mass stars. In Sect. \ref{anal_app}, 
we shall derive a mathematical expression for this statement. Note that the 
concentration of stars in the upper part of the distribution in Fig. 
\ref{model2_figs}a results from a higher density of individual SNe 
producing these abundance ratios (cf. the red curve in Fig. \ref{corr_12}e).
   
\par
  
In Model II, the relaxation of the instantaneous recycling approximation within
the star-forming regions introduces an evolutionary effect which has no 
counterpart in Model I. This effect should appear as a finite, non-vanishing 
dispersion in the different abundance ratios when the number of polluting SNe 
becomes large. The dispersion survives because stars that are formed early 
could only have been enriched by the most massive stars while stars that are 
formed late (i.e. $\sim 30~\mathrm{Myrs}$) could have been be enriched by SNe 
of any mass. However, the effect can not easily be detected in the the patterns 
in Fig. \ref{model2_figs}. In fact, this is as expected since we only use a 
maximum number of $20$ SNe in our simulations, which implies that the 
statistical fluctuations in the abundance ratios is still comparable to this 
dispersion. Eventually, the interaction between different clouds mixes the gas 
and the chemical abundance scatter in the subsequent generation of stars is 
decreased.

\par

\subsubsection{Possible sources of contamination}
\noindent
In our models we have assumed that the only stars that enrich a star-forming 
region with heavy elements have masses $\ge 10~\mathcal{M_{\odot}}$, excluding 
the very massive stars ($>100~\mathcal{M_{\odot}}$). This is consistent with 
the life-time of the star-forming regions adopted, i.e. $30$ Myrs. One could 
ask whether also intermediate-mass stars could be able to enrich a cloud. 
This would require that that either the life-time of the region is longer 
than assumed, or that the intermediate-mass stars were formed in epochs 
before the formation of the star-forming region. In the latter case it is reasonable to assume that the region has also been polluted by more 
short-lived, high-mass SNe. That is, none of the Halo stars would be enriched 
by a single, intermediate-mass star. The chemical patterns would be blurred but 
probably not very significantly. 

\par

The patterns could also possibly be affected by a population of very 
massive stars, preceding the onset of normal core collapse SNe. Similarly to 
the intermediate-mass stars, the very massive stars would pollute the 
star-forming regions with unknown amounts of elements, not accounted for 
in the models.

\par

Another, related source to noise affecting the patterns is the 
possible existence of stars sampling the nucleosynthetic signature of 
thermonuclear supernovae, like SNe type Ia. They are a different type 
of objects, not parametrized by the progenitor mass, and they will introduce a 
pattern in the $A/$ diagrams which is different from that of the core collapse 
supernovae. Progenitors of early thermonuclear SNe are thought to be close 
binary systems in which a white dwarf accrets matter from a subgiant star with
a mass of $2$-$3.5~\mathcal{M_{\odot}}$ (Branch 1998\nocite{branch98}). The
time delay between the onset of star formation and the formation of these 
thermonuclear SNe are at least on the order of $0.1$ Gyrs. Thus, the core 
collapse SN patterns in the $A/$ diagrams may not be severely contaminated by 
these objects. 

\par
   
As mentioned in Sect. \ref{assumptions} we do not consider metallicity 
dependent yields in our models. Such yields may also produce a smearing of the 
chemical patterns. The smearing is small for primary elements but could be as 
large as $0.5$ dex for secondary elements, e.g., for the ratio $^{14}\mathrm{N}/^{24}\mathrm{Mg}$ (Umeda et al. 2000\nocite{umeda00}).  

\par

In general, it would not be easy to recognize and remove stars 
enriched by these extra, hypothetical sources, except perhaps for stars 
with abundance patterns indicative of a pure thermonuclear SN contribution.
However, we have presently no strong reasons to believe that any of these 
sources have a significant effect on the $A/$ diagrams.

\subsubsection{The effects of statistical dependence}
\label{stat_indep}
\noindent  
Since the number of star-forming regions in reality is not infinite there is 
a certain probability that two, randomly picked Halo stars may have formed in 
the same cloud, or equivalently, the observed sample of stars may not be
completely statistically independent. If we randomly pick a number of stars 
which have been formed in a certain number of star-forming regions, how many of 
these regions are then represented by these stars? If the total mass 
of the Halo at the early epochs was ten times the mass of the stellar 
component today, which is $\sim 10^8~\mathcal{M_{\odot}}$ 
(Binney \& Merrifield 1998\nocite{bm98}), and all the gas was confined in 
star-forming regions of $10^6~\mathcal{M_{\odot}}$, there were approximately 
$1000$ different regions. Assume that every region formed an equal amount of 
stars. From available statistics (Christlieb \& Beers 2000\nocite{cb00} 
and references therein) we estimate that there are approximately $50~000$ extremely metal-poor Halo stars with a $B$ magnitude $\le 17$. Now, suppose 
that we randomly select a subsample of $100$ stars. How many different 
star-forming regions (sampling different chemical series) are then represented 
by the sample? Given the much larger number of available Halo stars, we 
estimate that approximately $95\%$ of the selected stars would sample different 
regions. On the other hand, if no more than $100$ star-forming regions existed 
in the early Galaxy, about $63$\% of the stars would still be statistically 
independent.  

\par

We performed a small test to investigate the necessity of the assumption of 
statistical independence. We generated two samples containing $1000$ stars 
each. The stars in the first sample were selected from individual chemical 
series where no star was enriched by more than $25$ SNe. These stars are 
statistically independent. The second sample consisted of $40$ chemical 
tracks, i.e., all stars enriched by $1-25$ SNe were selected from $40$ 
different chemical series. Stars belonging to a chemical track have a common chemical history and are statistically dependent. The difference between the 
two corresponding $A/$ diagrams was found to be relatively small and the 
characteristic pattern displayed by the first sample was well reproduced by the 
second one. Since only some tens independent chemical series contain enough 
information to form reliable patterns (remember that $100$ randomly selected 
Halo stars would sample maybe twice as many series), we conclude that the 
assumption of statistical independence is not vital for our conclusions.

\section{The origin of the chemical patterns}
\label{math}
\noindent
In this section we shall discuss the origin of the chemical patterns
found in the simulations above. We start with an approximate approach
to elucidate the source of the patterns, and also give some further
insight into the reason for the variety of patterns, demonstrated
in Fig. \ref{corr_diagrams} before we turn to the analytical theory.

\subsection{Approximate approaches to pattern formation}
\label{qualpf}
\noindent
Let us, to be explicit, assume that the yields for three elements, $A$, $B$ and
$C$, vary with mass such that, i.e, element $A$ is predominantly 
produced by SNe in a certain progenitor mass range, $m_{A'}$ to $m_{A''}$, 
with yield $p_{A'}$, and for the rest of the mass interval the yield is 
assumed to be much lower and constant, $p_{A''}$. Similarly, the element $B$ is 
assumed to be produced by stars in the disjoint mass range $m_{B'}$ to 
$m_{B''}$, with yield $p_{B'}$, while the stars outside the interval produce 
the element $B$ with the much lower yield $p_{B''}$. The element $C$ is assumed 
to be produced with a mass-independent yield $p_C$. Denote the number of SNe 
within the progenitor mass interval ($m_{A'}$,$m_{A''}$) by $n_1$, and 
correspondingly $n_2$ for the mass interval ($m_{B'}$,$m_{B''}$). The total 
number of SNe is denoted $n_3$. 

\begin{figure}
 \resizebox{\hsize}{!}{\includegraphics{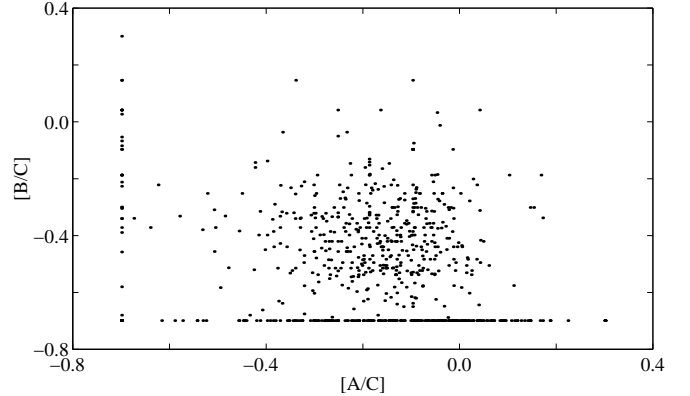}}
 \caption{Generic $A/A$ diagram generated from artificial yields. two   
  pronounced peaks in the $A$ and $B$ yields produce the characteristic
  "L"-shape with a cluster of stars round the centre of gravity}
 \label{artificial}
\end{figure}

\par

With these assumptions it is easy to derive the main properties of the 
distribution of stars in the [$A$/$C$]--[$B$/$C$] plane 
(Fig. \ref{artificial}). We find the total yields of element $A$, $B$ and 
$C$ to be, respectively

\begin{equation}
P_A = n_1p_{A'} + (n_3-n_1) p_{A''}
\end{equation}

\begin{equation}
P_B = n_2p_{B'} + (n_3-n_2) p_{B''}
\end{equation}

\begin{equation}
P_C = n_3p_C.
\end{equation}

\noindent
The extreme point in the lower left corner of the "L"-shaped distribution 
in Fig. \ref{artificial} will have the locus 
$\{\log(p_{A''}/p_C),\log(p_{B''}/p_C)\}$, and the end points of the "L" will 
be at $\{\log(p_{A'}/p_C),\log(p_{B''}/p_C)\}$ and 
$\{\log(p_{A''}/p_C),\log(p_{B'}/p_C)\}$, respectively. These points correspond 
to cases with no SNe in the mass intervals ($m_{A'}$,$m_{A''}$) {\it and} 
($m_{B'}$,$m_{B''}$), ($m_{B'}$,$m_{B''}$), and ($m_{A'}$,$m_{A''}$). 
Obviously, if these points can be observed, the yields can all be determined.

\par

If one further assumes that $n_1p_{A'} \gg (n_3-n_1)p_{A''}$ and that 
$n_2p_{B'} \gg (n_3-n_2)p_{B''}$ one finds the centre of gravity of the points 
in the $A/A$ diagram to be at 
$\{\log(p_{A'}/p_C) - \log(n_3/n_1),\log(p_{B'}/p_C) - \log(n_3/n_2)\}$. Thus, 
with the yields known observation of this point gives the ratios 

\begin{equation}
\frac{n_3}{n_1} = \frac{\int_{10}^{100} \phi(m)dm}{\int_{m_{A'}}^{m_{A''}} 
\phi(m)dm}
\end{equation}

\noindent
and

\begin{equation} 
\frac{n_3}{n_2} = \frac{\int_{10}^{100} \phi(m)dm}{\int_{m_{B'}}^{m_{B''}} 
\phi(m)dm}.
\end{equation}

\noindent
For the standard deviation in [$A$/$C$] and [$B$/$C$] one finds with the 
assumptions above $\sigma_{\mathrm{[}A/C\mathrm{]}}\sim 0.4/\sqrt{n_1}$ and 
similarly $\sigma_{\mathrm{[}B/C\mathrm{]}}\sim 0.4/\sqrt{n_2}$.

\par

These estimates are, however, very dependent on the assumed yields -- 
yields varying with stellar mass in more complex ways will contribute to
the standard deviations, and thus make determination of $n_1$ and $n_2$ in 
this way unrealistic.

\begin{figure*}[t]
 \resizebox{\hsize}{!}{\includegraphics{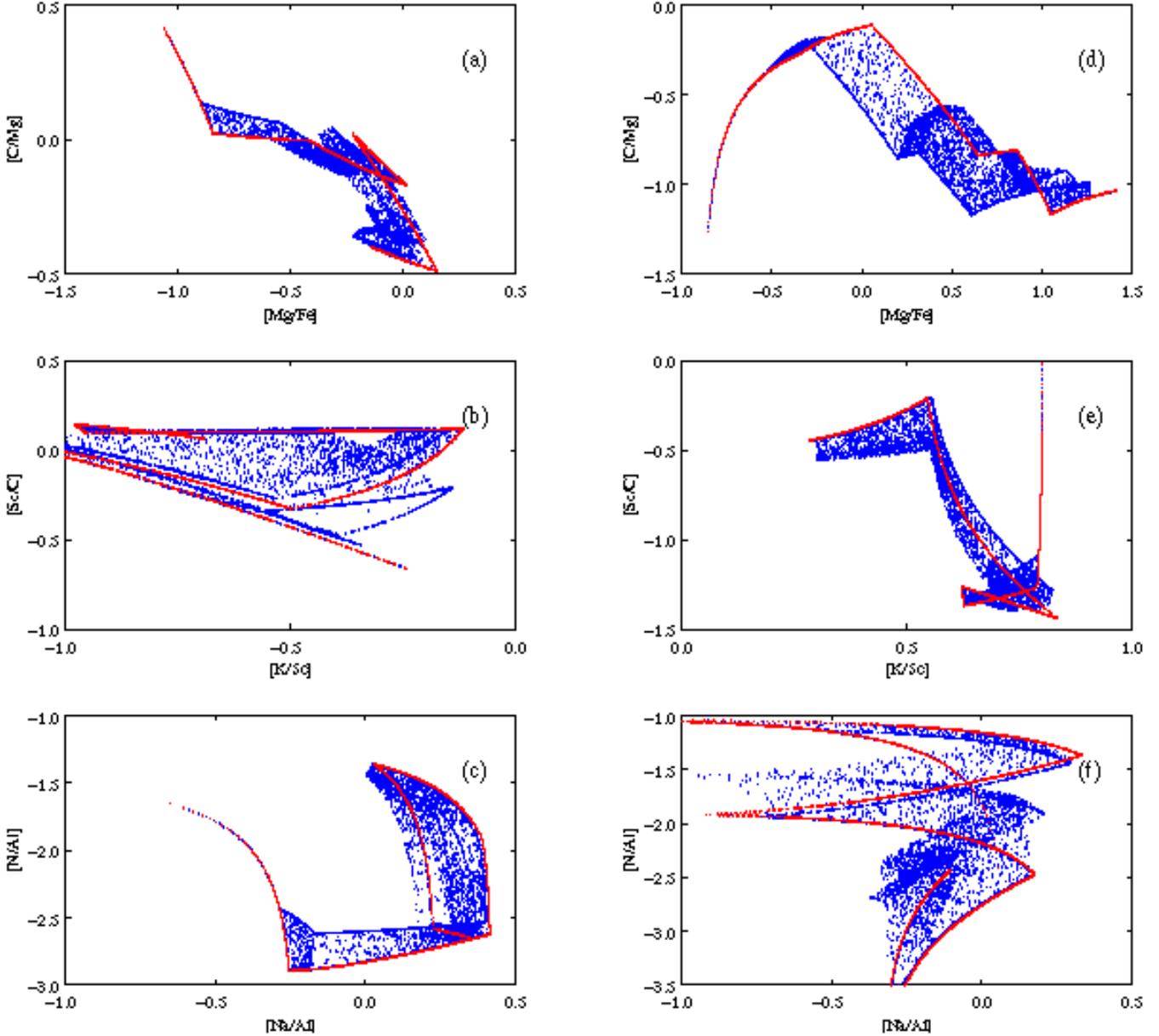}}
 \caption{As Fig. \ref{corr_diagrams} but for one and two polluting
  SNe only. The red dots denote stars enriched by one SN. These
  stars are located along one-dimensional curves in the abundance planes. The
  blue dots denote stars enriched by two SNe. Already two polluting
  SNe account for most of the structures present in the diagrams in 
  Fig. \ref{corr_diagrams}}
 \label{corr_12}
\end{figure*}

\par

The two legs of the "L"-shaped distribution in Fig. \ref{artificial} 
result from the SNe with no strong contribution of elements $A$ and $B$, 
respectively. The sparsely populated narrow sequences in Figs. 
\ref{corr_diagrams} and \ref{model2_figs} are, however, mainly the
result of the pollution of the star-forming region by just one SN. The
sequences are delineated by the range of SNe with different progenitor
mass. This is illustrated in Fig. \ref{corr_12}, where the corresponding 
distributions for just one SN (red dots), instead of $1-20$ SNe as in 
Fig. \ref{corr_diagrams}, have been plotted. If such narrow sequences could be 
identified observationally, which would require rich samples of very metal-poor
stars and high observational accuracy, one might directly read off the relative
yields at different progenitor masses. The values of the latter will, however, 
remain unknown. 

\par

In Fig. \ref{corr_12} we have also plotted the distributions of stars, 
resulting from cases with two polluting SNe (blue dots). It is clear 
from this figure, in comparison with Fig \ref{corr_diagrams}, that most of the 
structure of the $A/A$ diagrams is delineated already by models with two 
SNe. The density distribution in the diagrams of Fig \ref{corr_diagrams}, 
is, however, determined by the IMF as transformed by the mass dependence of the 
yields.

\subsection{An analytical approach}
\label{anal_app}
\noindent
We shall now proceed to a more exact treatment of the formation
of the chemical patterns by deriving analytical expressions for the density 
distribution (i.e. frequency distribution) of stars in the $A/$ diagrams. This 
distribution is represented by a two-dimensional density function. It is 
constructed by a sum of density functions, where each of these functions 
describes the distribution of stars enriched by a certain number of SNe. In 
order to understand these functions and get a feeling for the parameter 
dependence we shall begin by discussing one-dimensional density functions 
describing the distribution of a specific element or a ratio between two 
elements. The fundamental functions in this context are the ones that describe 
the distribution of stars enriched by individual SNe. In Appendix A we derive 
some general expressions for distributions of random variables and we shall 
frequently refer to those results in this section.

\par

The analytical theory is based on our simple model of chemical enrichment
(i.e. Model I). Its only important parameters are the stellar yields 
and the IMF slope index. The chemical patterns predicted by Model II are very 
much like the ones from Model I and the main result from Model I is not altered
even if some changes in the density distribution of stars occur. The 
conclusion is that the stellar yields, or rather the variations in the yields
with progenitor mass, play the crucial role for the shape of these density 
functions.

\subsubsection{The element distribution from individual supernovae}
\noindent
We start with deriving the fundamental expression for the distribution of stars
enriched by individual SNe. In general terms, the relative frequency of stars 
in an $A/$ diagram is given by the relative frequency of heavy element 
producing phenomena as a function of the amount of heavy elements produced. 
In this study, we assume that these elements are produced in SN explosions and 
that the number density of SNe is basically determined by $\phi$, the IMF. 
We have seen above that the actual number density is different from that of 
the IMF if, e.g., we consider continuous star-formation (Model II). However, 
this has not a big impact on the formation of the patterns.

\par

First, we need some definitions. The mass of a star formed in a 
star-forming region can be regarded as a random variable (r.v.) $M$, with a 
distribution function $F_{M}(m)$. The probability density function is then 

\begin{equation}
f_{M}(m) \equiv F_{M}'(m) \equiv \phi(m) = \phi_0 m^{-\alpha},
\label{fi}
\end{equation}

\noindent 
normalized as

\begin{equation}
\int\limits_{m_{\mathrm{min}}}^{m_{\mathrm{max}}}\phi(m) \mathrm{d}m = 1.
\end{equation}    

\noindent
We say that $M$ is an $\mathrm{IMF}(m_{\mathrm{min}},m_{\mathrm{max}})$-distributed 
random variable. Normally, the mass density, $m \phi(m)$, is normalized to
one but since we are interested in number densities we choose to normalize 
$\phi(m)$ instead. Furthermore, let the yield be a continuous function of 
stellar mass, $x=p(m)$, defined on the interval 
$[m_{\mathrm{min}},m_{\mathrm{max}}]$, i.e. $[10,100]$.  

\par

Now, as stellar masses are randomly distributed (according to this IMF) the 
amount of an element $A$, produced in a star of mass $m$, will also be randomly
distributed ($A$ stands for an arbitrary heavy element which
is a product of stellar nucleosynthesis). However, the distribution of the r.v.
$X_{(A)}=p_{(A)}(M)$ is different from that of $M$. The probability of
$X_{(A)}$ to be less or equal to $x$ is given by the distribution function 
$F_{(A)}(x)$, and according to Eq. (\ref{F_Y})

\begin{equation}
F_{(A)}(x) = F_M(p_{(A)}^{-1}(x)),
\label{cumul_func}
\end{equation}

\noindent  
assuming that $p_{(A)}(m)$ increases monotonically. For monotonically 
decreasing functions we will have that $F_{(A)}(x)=1-F_{M}(p_{(A)}^{-1}(x))$. 
We shall write the element in the subscript within parentheses to distinguish 
these functions from the final expression which is written without parentheses.

\begin{figure}
 \resizebox{\hsize}{!}{\includegraphics{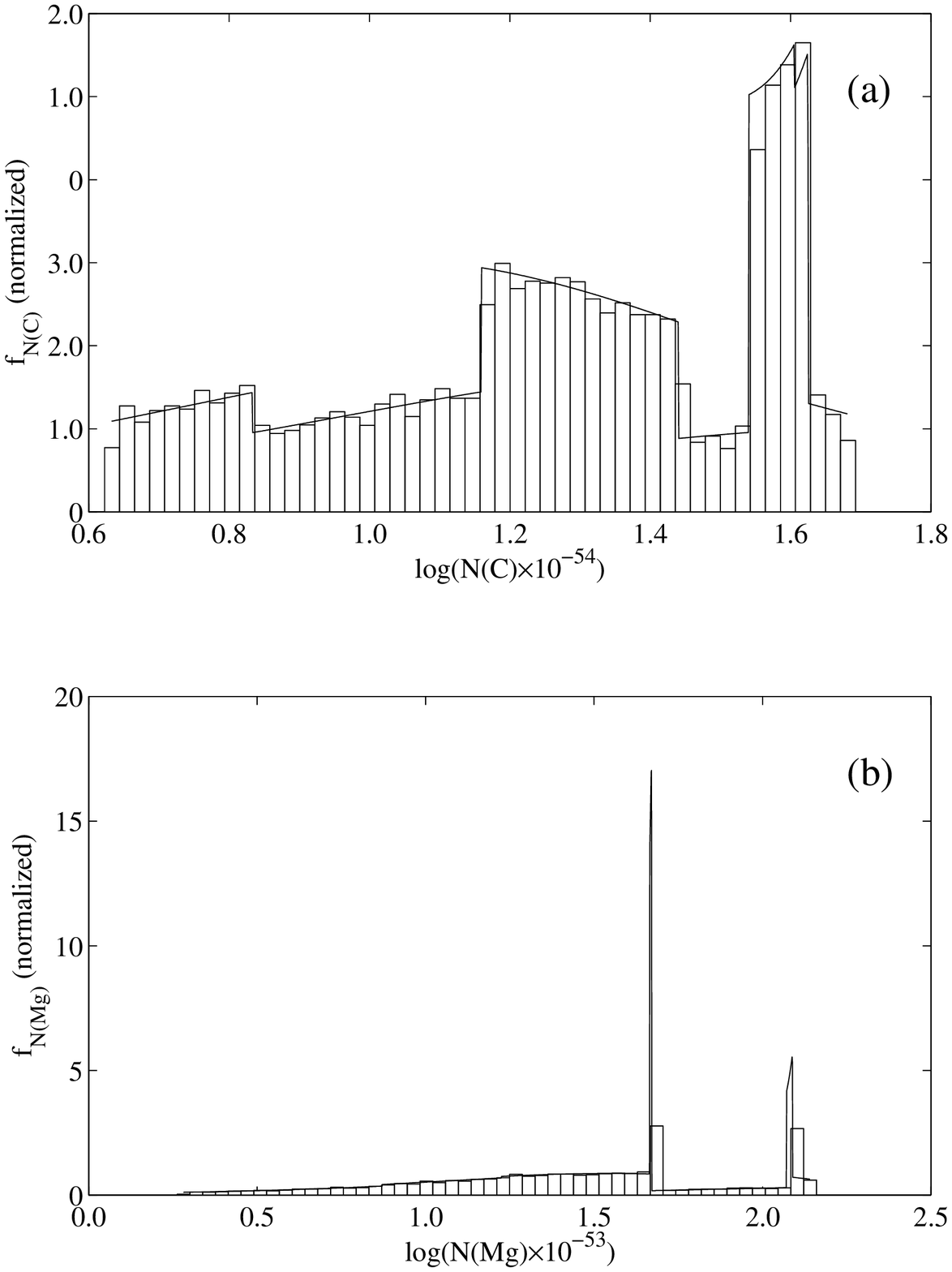}}
 \caption{\textbf{a)} The density function of carbon from individual SNe
  with the yield ($Z=0$ models) taken from WW95 (see Fig. \ref{basic}d). The
  Salpeter IMF is used. The histograms are binned data from numerical 
  simulations of $20~000$ clouds with a single SN explosion in each 
  cloud (Model I) while the full line is the analytical function calculated 
  from Eq. (\ref{f_X}) and transformed according to Eq. (\ref{transform}).
  \textbf{b)} The corresponding density function for the element magnesium}
 \label{f_X_fig}
\end{figure}

\par

As our aim is to derive expressions for the density of stars in the $A/$ 
diagrams, we shall describe the distributions of random variables in terms of 
density functions. The density function, $f_{(A)}(x)$, of $X_{(A)}$ is 
given by the derivative of $F_{(A)}(x)$ with respect to $x$. Eq. (\ref{f_Y})
together with Eq. (\ref{fi}) gives that

\begin{equation}
f_{(A)}(x) = \frac{\mathrm{d}}{\mathrm{d}x}F_{(A)}(x) = 
\left| \frac{1}{p_{(A)}'(p_{(A)}^{-1}(x))}\right| \times
\phi(p_{(A)}^{-1}(x)),
\label{f_X}
\end{equation} 

\noindent
where $p_{(A)}' \equiv \mathrm{d}p_{(A)}/\mathrm{d}m$ and $p_{(A)}^{-1}$ is the
inverse function to $p_{(A)}$. Note that 
$x \in \mathrm{[}X_{(A)}^{\mathrm{min}},X_{(A)}^{\mathrm{max}}\mathrm{]}$, i.e. 
$f_{(A)}(x)$ as well as $F_{(A)}(x)$ are defined on the interval between the
minimum and the maximum produced amount of element $A$ in stars with masses in 
the interval $m \in \mathrm{[}m_{\mathrm{min}},m_{\mathrm{max}}\mathrm{]}$.

\par

Eq. (\ref{f_X}) holds for monotonic yields, $p_{(A)}$. If the yield 
has local extrema there is no way of finding a single inverse to $p_{(A)}$. As
in Eq. (\ref{f_Y}) it is necessary to split the interval of $m$ such 
that the yield is monotonic on each subinterval. 

\par

Let us consider a simple example. Assume that an element $A$ is produced in 
massive stars in such a way that the stellar yield $x \equiv p_{(A)}(m) 
\propto m^{1/2}$ (measured in $\mathcal{M_{\odot}}$). Furthermore, let the 
high-mass stars be distributed according to an $\mathrm{IMF}\propto m^{-2}$. 
Thus, $p_{(A)}^{-1}(x) \propto x^2$ and $|p_{(A)}'(m)| \propto m^{-1/2}$. Using 
Eq. (\ref{f_X}) we find that the number distribution of massive stars 
producing a specific amount of element $A$ is proportional to 
$(x^2)^{-2}/(x^2)^{-1/2}=x^{-3}$. On a logarithmic scale the density
function $f_{(A)}(x) \propto 10^{-2x}$ according to Eq. (\ref{transform}).  

\par

The density function, $f_{(A)}(x)$, describes the relative distribution of 
stars enriched by individual SNe, where each SN produces a 
certain amount of each element. It consists of two factors. The second
factor, the IMF, accounts for the non-uniform distribution of stellar masses. 
It is a smooth, monotonically decreasing function of mass, which, in practice, 
means that it will not produce any sudden changes in $f_{(A)}(x)$. On the other 
hand, the first factor, which depends on the stellar yield, may change 
drastically with mass. This is then reflected in the density function. 
Fig. \ref{f_X_fig} shows two examples of density functions for the elements 
carbon and magnesium. The yields are given by the zero-metallicity models of 
WW95\nocite{ww95}. 

\par

Note that we have not yet been discussing distributions of stars (i.e. 
low-mass stars) enriched in heavy elements. Thus, the variable $x$ in 
$f_{(A)}(x)$ is not the amount of an element in a low-mass star. So far, $x$
represents the amount of the element produced in the SN that has 
enriched the low-mass star. In general, detailed knowledge about the mixing of 
the SN material with the ambient, possibly pre-enriched medium is needed
in order to determine $A$/H ratios. In our simulations, this is 
accounted for by assuming a constant mass of the hydrogen clouds and no global 
mixing and we shall adopt the same mixing scenario here. This questionable 
assumption is of small significance for the pattern formation in the $A/A$ 
diagrams, i.e. for elements beyond hydrogen.

\subsubsection{The distribution from two or more supernovae}
\label{sect_convolve}
\noindent
If more than one SN explodes in each cloud the distribution of 
the freshly synthesized material in the low-mass stars is no longer described 
by the density function $f_{(A)}(x)$. The sum of the contributions from
every SN has to be considered, which alters the distribution. For example, two 
different sets of SNe may well produce the same total amount of an 
element. Thus, the correct density function describes a sum of the 
independent, equally distributed random variables, $X_{(A)}$. This function is 
given by a convolution of $f_{(A)}(x)$ with itself.

\par

The convolution formula for two random variables (i.e. two SNe) is 
derived in Appendix A. Here, we shall give the general expression for $n$ 
random variables since we are interested in the shape of the density function
when we sum up the contributions from $n$ SNe. A generalization of 
Eq. (\ref{cf_04}) is done by adding a third r.v. to the first two, then adding 
a fourth and so on. Thus,

\begin{eqnarray}
f_{n(A)}(x) & = &
\int\limits_{-\infty}^{+\infty}...\int\limits_{-\infty}^{+\infty}
f_{(A)}(x_1)f_{(A)}(x_2-x_1)\times...\nonumber\\
& \times & 
f_{(A)}(x_{n-1}-x_{n-2})f_{(A)}(x-x_{n-1})\mathrm{d}x_1...\mathrm{d} x_{n-1}.
\label{f_nX}
\end{eqnarray}

\noindent 
As for the expression in Eq. (\ref{cf_04}) the integrals are taken over the 
whole space. The function describes the distribution of the sum of $n$ 
independent, equally distributed r.v.s $X_{(A)}$. This is indicated by the 
notation $n(A)$ in the subscript of $f_{n(A)}(x)$.

\begin{figure}
 \resizebox{\hsize}{!}{\includegraphics{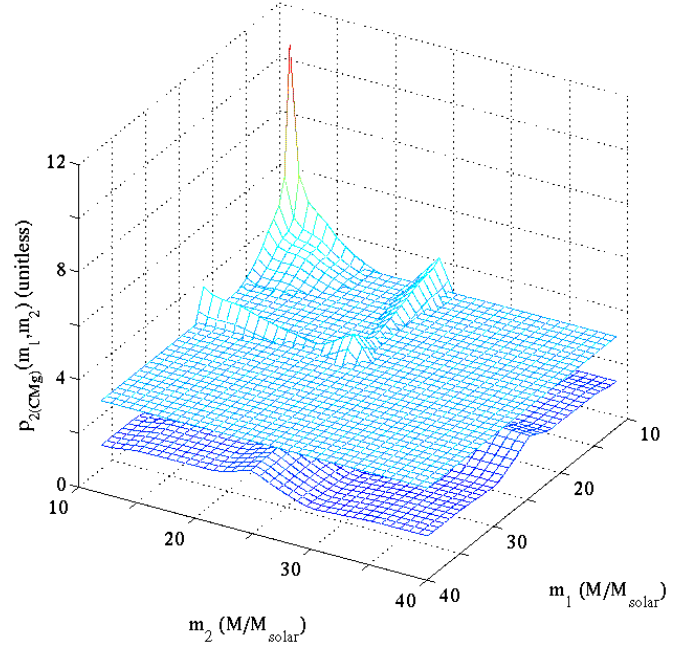}}
 \caption{The two-dimensional yield ratio C/Mg as given by Eq. (\ref{p_ratio}).
  Also shown in the figure is the plane at $p_{2\mathrm{(CMg)}}=3.50$. 
  Every plane parallel to the $m_1m_2$-plane which intersects the yield 
  function at a certain point defines an integration region for which the 
  integral in Eq. (\ref{F_nXratio}) can be calculated, giving the value of the 
  distribution function in the point $x = p_{2\mathrm{(CMg)}}$}
 \label{pCMg_2D}
\end{figure}

\par

It is worth mentioning a couple of properties of the convolution described
by Eq. (\ref{f_nX}). When $n$ tends to infinity the corresponding density
function of the arithmetical mean, i.e. the function derived from 
Eq. (\ref{mean}), tends to a Dirac delta function centred at 
$x=\int p(m) \phi(m) \mathrm{d}m$. Moreover, a sum of random variables can be 
approximated by a normal distribution according to the Central Limit Theorem. 
The dispersion of the arithmetical mean is then proportional to $n^{-1/2}$.

\par

The number of SNe needed for this approximation to be valid depends 
ultimately on the shape of $f_{(A)}(x)$, but $n \sim 20$ is probably a good 
order-of-magnitude estimate. The averaging gradually erases the specific 
structures in the density functions and as soon as the shape of $f_{n(A)}(x)$ 
becomes well-behaved (i.e. Gaussian) it does not carry any information about
the original yield and no new patterns are formed in the $A/$ diagrams (see, 
e.g., the decreasing scatter at the high-metallicity end in Fig. \ref{CMg_fig}b
or the crowding of stars round the point ($-0.19,-0.25$) in 
Fig. \ref{corr_diagrams}a).

\subsubsection{Abundance ratios}
\noindent
In this section we shall derive the important expression for the distribution 
of an abundance ratio between two heavy elements, $A$ and $B$.  

\par

Let us first define a generalized stellar yield. The total amount of an 
element ejected by $n$ SNe can be regarded as a yield in the 
$n$-dimensional $m$-space, i.e.

\begin{equation}
p(m_1,...,m_n)=\sum\limits_{i=1}^n p(m_i),
\end{equation}

\noindent
where $p(m)$ is the normal, one-dimensional stellar yield for a star of mass
$m$. In this way the yield ratio of two elements is simply defined as

\begin{equation}
p_{n(AB)}(m_1,...,m_n)=\frac{\sum\limits_{i=1}^n 
p_{(A)}(m_i)}{\sum\limits_{i=1}^n p_{(B)}(m_i)}.
\label{p_ratio}
\end{equation}

\noindent
As before, the subscript $n$ stands for the number of polluting SNe 
which in this case is equal to the number of dimensions as the original stellar
yield is one-dimensional.

\begin{figure}
 \resizebox{\hsize}{!}{\includegraphics{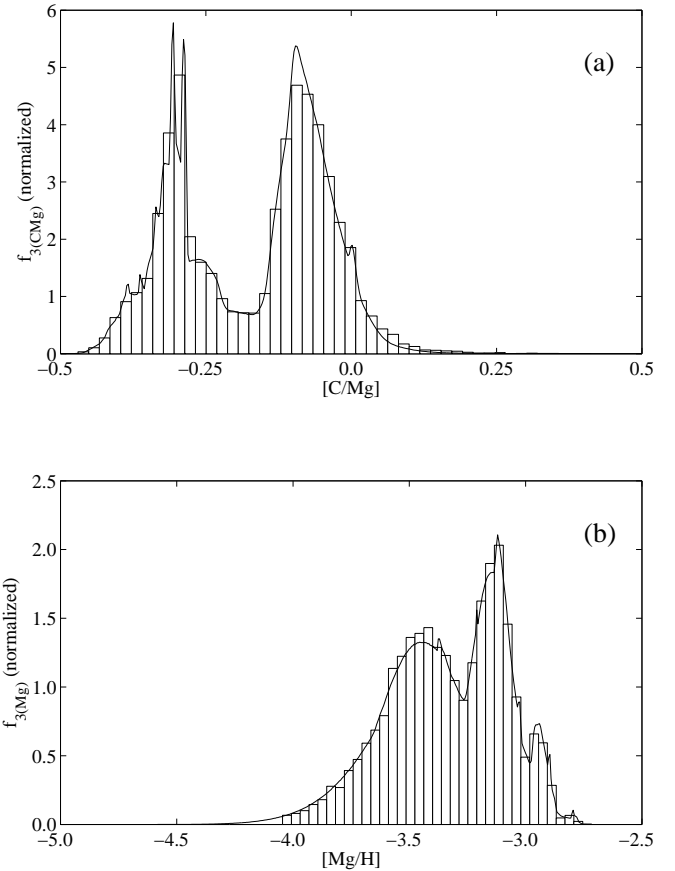}}
 \caption{\textbf{a)} The density function of low mass-stars in [C/Mg]. 
  Each star has been enriched by 3 SNe. Their masses were distributed 
  according to the Salpeter IMF. As before, the histogram shows binned data 
  from a numerical simulation while the full line is the corresponding density 
  function to the formal solution of the distribution function in 
  Eq. (\ref{F_nXratio}). \textbf{b)} The density function for [Mg/H]. This is 
  in principle a convolution of the random variable $X_{\mathrm{(Mg)}}$ (see
  Fig. \ref{f_X_fig}b) using Eq. (\ref{f_nX}) for $n=3$. The magnesium 
  abundance is measured relative to $5\times10^{5}~\mathcal{M_{\odot}}$ of 
  hydrogen}
 \label{f_CMg_fig}
\end{figure}

\par

Now, similarly to the multi-dimensional random variable in Eq. (\ref{app_01}), 
we form the r.v. $X_{n(AB)}=p_{n(AB)}(M_1,...,M_n)$. This ratio has properties 
similar to that of the mean of a single element $A$. It is possible to show 
that the distribution function $F_{n(AB)}(x)$ can be written as

\begin{equation}
F_{n(AB)}(x) = \int...\int_{D_{x}} \phi(m_1) \times...
\times \phi(m_n)\mathrm{d}m_1...\mathrm{d}m_n
\label{F_nXratio}
\end{equation}

\noindent
according to Eq. (\ref{app_02}). The integration region $D_x$ is defined by 
the equation

\begin{equation}
p_{n(AB)}(m_1,...,m_n) \le x,
\label{D_x}
\end{equation}

\noindent
i.e. the part of the $m_1...m_n$-hyper plane where the yield is less than or
equal to $x$ (see Fig. \ref{pCMg_2D}). Note that because all stellar mass 
r.v.s are independent the integrand in Eq. (\ref{app_02}) reduces to a simple 
multiplication of the IMFs. 

\par

The integrand in Eq. (\ref{F_nXratio}) is almost trivial to handle while the
integration region is highly non-trivial. From Eq. (\ref{D_x}) we see that 
depending on how the yield varies, $D_x$ may not even be simply connected.
This is illustrated in Fig. \ref{pCMg_2D}. The integral should be calculated 
for every $x$ which then gives the distribution function. The density 
function is obtained taking the derivative with respect to $x$ as in 
Eq. (\ref{f_X}). Fig. \ref{f_CMg_fig} shows two examples of density 
functions of such compound random variables. Fig. \ref{f_CMg_fig}a shows the 
distribution of stars over [C/Mg] for three polluting SNe using the same 
yields as above. The full line is a solution to Eq. (\ref{F_nXratio}). The 
density function in Fig. \ref{f_CMg_fig}b is calculated from Eq. (\ref{f_nX}). 
Both functions are transformed to a logarithmic scale (and translated 
relative to solar values) using Eq. (\ref{transform}).

\par

\subsubsection{The construction of $A/$ diagrams}
\label{aadiag}
\noindent
We are ultimately interested in two-dimensional distributions describing the
density of extremely metal-poor low-mass stars in the $A/$ diagrams. If two 
random variables are independent their joint density equals the product of
the individual densities (see the derivation of the convolution formula in 
Appendix A where independence is assumed). However, the random variables we 
discuss here are dependent. If we simultaneously observe the 
random variables in Fig. \ref{f_CMg_fig}a and b the corresponding density 
function is not a multiplication of the two individual density functions. 
It is not possible to obtain every value of the ratio [C/Mg] for a given
value of [Mg/H] by the combination of three SNe as the two variables are 
entangled via the progenitor masses of these SNe.

\begin{figure}
 \resizebox{\hsize}{!}{\includegraphics{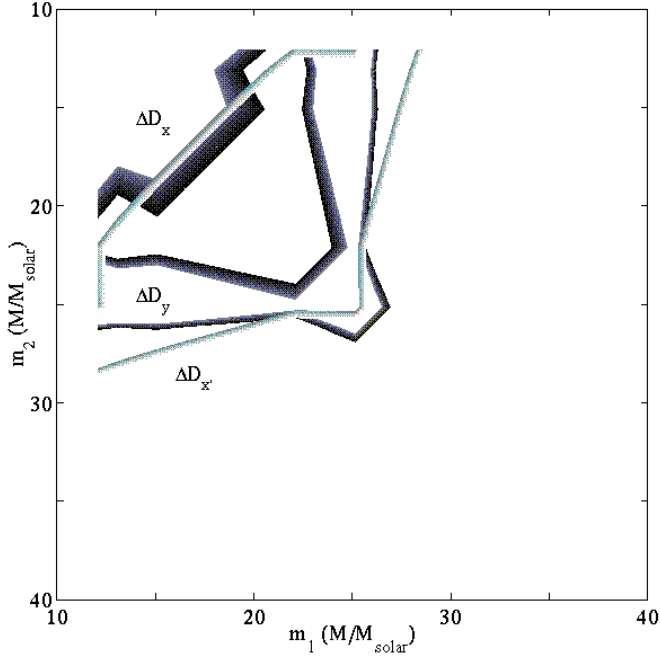}}
 \caption{Intersecting integration regions. $\Delta D_y$ (dark-shaded areas)
  is defined by $p_{2\mathrm{(CMg)}}=y\sim3.50$ (cf. Fig. \ref{pCMg_2D}) 
  while $\Delta D_x$ and $\Delta D_{x'}$ (light-shaded areas) are defined by
  $p_{2\mathrm{(Mg)}}=x\sim 0.10$ and $p_{2\mathrm{(Mg)}}=x'\sim 0.20$ 
  respectively. Hence, $\Delta D_{xy}$ and $\Delta D_{x'y}$ are given by the
  intersections $\Delta D_x \bigcap \Delta D_y$ and $\Delta D_{x'} \bigcap 
  \Delta D_y$. A calculation of the integral in Eq. (\ref{F_nXY}) over 
  $\Delta D_{xy}$ for every $x=p_{2\mathrm{(Mg)}}(m_1,m_2)$ and 
  $y=p_{2\mathrm{(CMg)}}(m_1,m_2)$ gives the two-dimensional density 
  function $f_{2(\mathrm{Mg},\mathrm{CMg})}$}
 \label{deltaD}
\end{figure}

\par

Suppose that we have four elements $A$, $B$, $C$, and $D$, where not all 
elements are necessarily different. Now, the joint distribution of $X_{n(AB)}$ 
and $X_{n(CD)}$ (or, e.g., $X_{n(A)}$ and $X_{n(BC)}$) is described by the 
function $f_{n(AB,CD)}$ (or$f_{n(A,BC)}$). As usual, the subscript $n$ stands 
for the number of contributing SNe. The derivation is similar to that of 
the one-dimensional distribution function in Eq. (\ref{F_nXratio}). Here, the 
density functions are calculated directly, as in Eq. (\ref{app_07}). Thus,

\begin{eqnarray}
f_{n(AB,CD)}(x,y) & = & \int...\int_{\Delta D_{xy}} \phi(m_1) 
\times... \nonumber\\
& \times & \phi(m_n)\mathrm{d}m_1...\mathrm{d}m_n,
\label{F_nXY}
\end{eqnarray}

\noindent
where the integration region, $\Delta D_{xy}$ is defined via the equations

\begin{equation}
\left\{ \begin{array}{ll}
\Delta D_{x}:~x<p_{n(AB)}(m_1,...,m_n)\le x+\mathrm{d}x\\
\Delta D_{y}:~y<p_{n(CD)}(m_1,...,m_n)\le y+\mathrm{d}y.
\end{array} \right.
\end{equation}

\noindent
Thus, for every point $(x,y)$ the integration is performed over the small
common region in $m$-space defined by the intersection, 
$\Delta D_{x} \bigcap \Delta D_{y}$. The density of stars (enriched by 
$n$ SNe) in the [$A$/H]--[$B$/$C$] plane, described by $f_{n(A,BC)}(x,y)$, 
is calculated similarly. However, the integration region is changed to 
(see Fig. \ref{deltaD}) 

\begin{equation}
\left\{ \begin{array}{ll}
\Delta D_{x}:~x<p_{n(A)}(m_1,...,m_n)\le x+\mathrm{d}x\\
\Delta D_{y}:~y<p_{n(BC)}(m_1,...,m_n)\le y+\mathrm{d}y.
\end{array} \right.
\end{equation}

The integrands in Eq. (\ref{F_nXratio}) and Eq. (\ref{F_nXY}) are the same. 
The difference lies again in the integration regions and we note that the 
(generalized) stellar yields are important components in the expression for 
the abundance ratio as the functions defining these regions. Moreover, the IMF 
only partly determines the density in each point, together with the gradient of 
the yield function. This is clearly seen in Eq. (\ref{f_X}) but can also be 
traced in Fig. \ref{deltaD} where the differential areas have different 
sizes depending on the shape of the yields.

\begin{figure}
 \resizebox{\hsize}{!}{\includegraphics{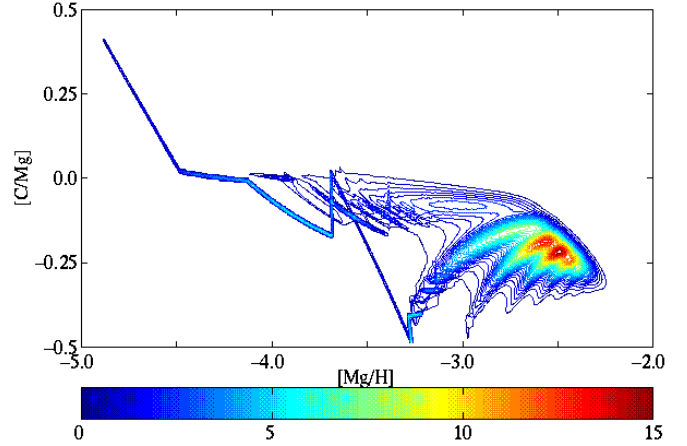}}
 \caption{Contour plot of $f_{\mathrm{Mg},\mathrm{CMg}}$ 
  (cf. Fig. \ref{CMg_fig}a) computed for 1--20 SNe with $w_i=1/20=0.05$ 
  using the yields from WW95. The IMF is of Salpeter form. The relative 
  number density of stars is indicated by the colour bar}
 \label{densf_CMg}
\end{figure}

\begin{figure}
 \resizebox{\hsize}{!}{\includegraphics{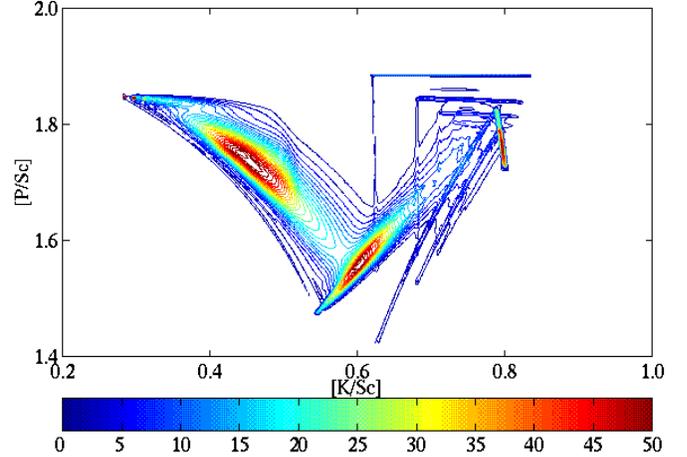}}
 \caption{Contour plot of $f_{\mathrm{KSc},\mathrm{PSc}}$. The model 
  parameters are the same as in Fig. \ref{densf_CMg}. The stellar yields are
  taken from Netal97. The colour bar indicates the relative number density 
  of stars}
 \label{densf_KPSc}
\end{figure}

\par

The final step to an analytical expression for the distribution of stars in an 
$A/$ diagram is a summation of the density functions over the number of SNe. 
The expression for the density in an $A/A$ diagram, i.e. the density in 
the [$A$/$B$]--[$C$/$D$] plane, is 

\begin{equation}
f_{AB,CD}(x,y)=\sum\limits_{i=1}^{n} w_i \times f_{i(AB,CD)}(x,y).
\label{finalexpr}
\end{equation}

\noindent
The weight $w_i$ is the fraction of clouds forming $i$ SNe. If we substitute 
$f_{i(A,BC)}(x,y)$ in the sum, we instead obtain a function formally describing 
the density of stars in the [$A$/H]--[$B$/$C$] plane. 
Note that the first term in the sum is a one-dimensional function since any 
star enriched by a single SN has a unique chemical composition determined by 
the yields of that SN. Thus, the density function maps the one-dimensional 
yield ratio on the $A/A$ plane. The function is seen as a curve in 
Figs. \ref{densf_CMg} and \ref{densf_KPSc} (see also Fig. \ref{corr_12}). The
form of this curve actually determines the form of the total density 
function to a large extent.

\par

We end this section by a small note on scale transformations.
The appropriate equations are derived in Appendix A. In our derivations all 
functions are on a linear scale. Often astrophysical quantities may span 
several orders of magnitude. It is then more advantageous to display these 
quantities on a logarithmic scale. The transformation is made using Eq. 
(\ref{transform}) and Eq. (\ref{transform2D}) for one- and two-dimensional 
functions respectively. However, it is also possible to directly take the 
logarithm of the yield or the generalized yield and use these functions in 
the expressions for the densities instead.

\subsection{The robustness of the $A/A$ diagrams}
\label{ratio}
\noindent
It is not probable that patterns like those in Fig. \ref{densf_CMg} can be 
observed, since they are based on oversimplified models. The star-forming 
regions have different masses. Mixing within the regions is not complete and 
global mixing of remnant, enriched gas, or infalling gas, and the formation 
of a second generation of star-forming regions probably occurred even for 
the extremely metal-poor Halo. By neglecting all these effects the 
treatment of the mixing in our models is oversimplified 
(see Nakasato \& Shigeyama 2000\nocite{nakasato00}). Thus, a horizontal 
smearing effect occurs in $A/$H diagrams such as [Mg/H]--[C/Mg], and the 
patterns are most likely lost in these diagrams. 

\begin{figure}
 \resizebox{\hsize}{!}{\includegraphics{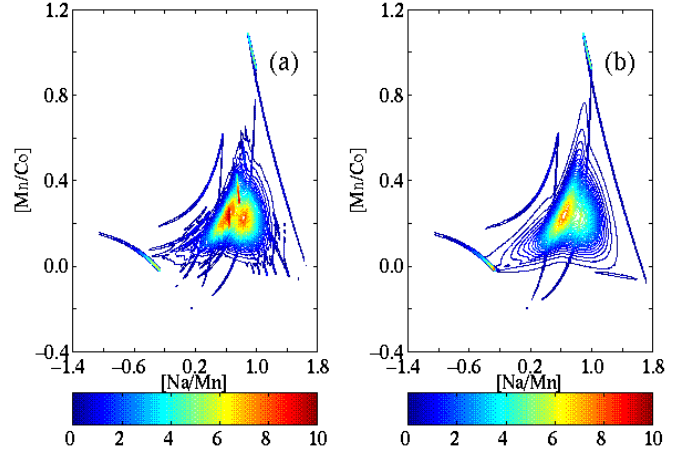}}
 \caption{Contour plots of the density of stars in the [Na/Mn]--[Mn/Co]
  plane with yields from Netal97. \textbf{a)} The density function with 
  unweighted yield terms (i.e. $\xi_i=1$ for all $i$). \textbf{b)} The 
  corresponding function with weighted yield terms according to 
  Eq. (\ref{p_ratio_mod}) representing 
  the effect of global mixing. The weights have a uniform distribution, ranging 
  from $0$ to $1$. Due to the weights the microstructure in the second $A/A$ 
  diagram has been wiped out. Otherwise, the difference between the two 
  diagrams is very small}
 \label{NaMnCo_diff}
\end{figure}

\par

The patterns in the $A/A$ diagrams are, however, not
sensitive to the masses of the star-forming regions, i.e. the amount of
ambient gas that is mixed with the SN ejecta. If the SN remnant material is 
mixed with a pre-enriched ISM one might yet think that the information on the 
original production sites of the elements can not be traced in the observed 
abundance ratios, as different fractions of the enriched medium may contribute 
differently to different stars. This occurs when a second generation 
of star-forming regions is formed. Suppose for example that a new 
region is formed out of the dispersed gas of two former such regions with the 
proportions $1:3$, i.e. $25\%$ of the total mass originates from the first 
region and $75\%$ of the mass from the second region. Assume further that 
these former regions were enriched by one SN each and the new region 
produces one SN. An abundance ratio in the gas and in 
subsequently formed stars is then found by summing up the individual 
contributions from each SN, as in Eq. (\ref{p_ratio}). However, each 
term has a a weight associated with it, depending on how great the contribution 
is from each region. In our example the weights are $0.25,~0.75$ 
and $1.00$. Generally, Eq. (\ref{p_ratio}) has to be modified such that

\begin{equation}
p^{\xi}_{n(AB)}(m_1,...,m_n)=\frac{\sum\limits_{i=1}^n \xi_i 
p_{(A)}(m_i)}{\sum\limits_{i=1}^n \xi_i p_{(B)}(m_i)}.
\label{p_ratio_mod}
\end{equation}

\noindent
The only difference between this expression and the one in Eq. (\ref{p_ratio})
is the weight $\xi_i$ on each yield. The weight is supposed to mimic the effect
of large-scale mixing and turbulent motions in the ISM and can be regarded 
as a random variable similar to the mass such that the r.v. $X^{\xi}_{n(AB)}$ 
is

\begin{equation}
X^{\xi}_{n(AB)}=\frac{\sum\limits_{i=1}^{n}\Xi_i p_{(A)}(M_i)}{\sum\limits_{i=1}^{n}\Xi_i
p_{(B)}(M_i)}=p^{\xi}_{n(AB)}(\Xi_1,M_1,...,\Xi_n,M_n)
\label{X_ratio_mod}
\end{equation}

\noindent
where the superscript $\xi$ denotes the fractional contribution defined by the 
weighted generalized yields in Eq. (\ref{p_ratio_mod}). By introducing these
weights, the generalized yield $p_{n(AB)}(m_1,...,m_n)$ looses its symmetry 
(e.g. in Fig. \ref{pCMg_2D}, the two-dimensional yield ratio would no longer be 
symmetric around $m_1=m_2$). Thus, depending on the weights, a specific 
abundance ratio may be produced by several different sets of SNe. This 
means that if we have no information on the weights we are no longer able to
identify the set of SNe that gives the observed ratio. 

\par

Even so, simulations with randomly distributed weights show that the 
distribution of stars in $A/A$ diagrams is practically 
unaltered (see Fig. \ref{NaMnCo_diff}) with respect to the original 
distribution for which all $\xi_i=1$. This is due to the fact that the 
intersection between the two regions $\Delta D_x^{\xi}$ and 
$\Delta D_y^{\xi}$ is non-zero (thus, the 
integral in Eq. (\ref{F_nXY}) is non-zero) in almost the same region in yield 
space (i.e. abundance ratio space) as for the unperturbed (unweighted) 
integration region $\Delta D_{xy}$. Note however, that
the integration region over stellar masses may be different and far from
symmetric which changes the value of the density function in each 
point. This is also observed in Fig. \ref{NaMnCo_diff}. The distributions 
have the same shape but the density of stars in the structures is 
somewhat different. Furthermore, in Fig. \ref{NaMnCo_diff}b the microstructure 
is wiped out. It is clear that some information must be lost by 
introducing weights on the yield terms. Thus, we have no longer knowledge of
the absolute amounts of the elements produced in the SNe. 

\par

Even though the two chemical patterns in Fig. \ref{NaMnCo_diff} look 
qualitatively similar a relevant question is whether the microstructure in 
Fig. \ref{NaMnCo_diff}a is crucial for a reconstruction of SN yields. This 
structure originates from the one-dimensional curve sampling individual SNe 
(cf. the red curves in the diagrams of Fig. \ref{corr_12}) or sooner it 
corresponds to this curve for the case $n=2,3,$ etc. In principle, this structure does not contain any unique information and the loss of it 
should not be vital for a yield reconstruction procedure. In this particular 
example the pattern in Fig. \ref{NaMnCo_diff}a also shows two distinct 
peaks of which the right one is the convergence point or the centre of gravity. 
This dichotomy is not as prominent in Fig. \ref{NaMnCo_diff}b. Would the 
lack of a second peak lead to a significantly different result when 
reconstructing the yields? We believe that this is not the case. A comparison 
with the much greater diversity of patterns shown in Fig. \ref{corr_diagrams} 
suggests that the major part of the information content on the yields in 
Fig. \ref{NaMnCo_diff}a still remains in Fig. \ref{NaMnCo_diff}b. A number of 
similar simulations for other elements confirm this result. However, more 
tests should be carried out in order to quantitatively determine the 
sensitivity of the yields on the details of the chemical patterns.

\section{Discussion -- Can patterns be observed?}
\label{discussion}
\noindent
Above, we have discussed the formation of patterns in $A/$ diagrams and the 
possibilities of using abundance ratios as a diagnostic of early stellar 
nucleosynthesis. However, the question remains: Are chemical abundance 
patterns detectable in practice? 

\par 

\begin{figure}
 \resizebox{\hsize}{!}{\includegraphics{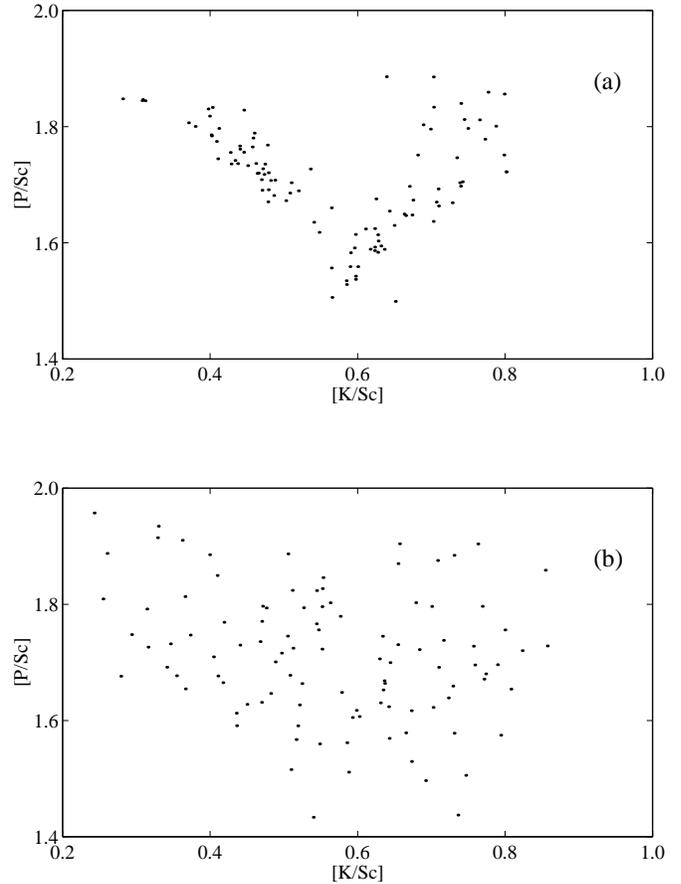}}
 \caption{\textbf{a)} A scatter plot of around $100$ stars (enriched by 
  $1$--$15$ SNe) in the [K/Sc]--[P/SC] plane. The nucleosynthesis calculations 
  by Netal97 have been used. The "V"-shape is easily detected (cf. Fig. 
  \ref{densf_KPSc}). \textbf{b)} The corresponding distribution with an 
  additional uncertainty of $0.08$ dex in the abundance ratios. The "V"-shape 
  is less pronounced but still detectable}
 \label{KPSc_cf}
\end{figure}

It is difficult to estimate the accuracy needed in the abundance determinations
and the minimum number or sample stars that is required for detecting the 
chemical patterns. Since the number of Halo stars that can be observed 
accurately is limited, these parameters are dependent. Fig. \ref{KPSc_cf} shows
a comparison between the original distribution of about $100$ stars in the 
[K/Sc]-[P/Sc] plane (cf. Fig. \ref{densf_KPSc}) and the corresponding 
distribution for which an uncertainty of $0.08$ dex is added in the observed
abundance ratios. The original "V"-shape is barely detected in 
Fig. \ref{KPSc_cf}b. However, the pattern would more easily appear if we could 
double the number of stars. Larger uncertainties in the abundance ratios would 
completely wipe out the pattern and this can not be compensated for by adding 
more stars. On the other hand, the "U"-shape in Fig. \ref{corr_diagrams}c is 
detectable even for uncertainties above $0.1$ dex in the abundance ratios. 
However, for making that pattern visible at all several hundred stars must be 
observed, many more than are needed for the "V"-shaped pattern. Thus, 
depending on the shape, some patterns are more sensitive to observational 
uncertainties than to the number of sample stars while the opposite is true for 
other types of patterns.
 
\par

This discussion also holds for $A/$H diagrams (e.g. Fig. \ref{CMg_fig}). 
Assuming that there are no intrinsic uncertainties which 
wipe out the patterns, we estimate, by convolving the density function in 
Fig. \ref{densf_CMg} with a Gaussian profile, that the structures become 
undetectable if the uncertainties in the derived abundances are larger than 
$\sim 0.05$ dex. This is an upper limit as we in reality only deal with a 
limited number of sample stars, i.e. poorer statistics. However, e.g., the 
bimodal structure in the [N/O]--[O/H] plane (see Fig. 3 in 
Karlsson \& Gustafsson 2000) caused by a pronounced peak in the N-yield 
(WW95\nocite{ww95}) is not erased even for uncertainties in 
the abundances as large as $0.15$ dex. This also suggests that large-scale 
structures in $A/$H diagrams can possibly survive the effect of mixing in 
the interstellar medium, if the mass range of typical star-formation regions
is limited to within a factor of $2$ or so. One should also note that similar
patterns could also be expected to be visible in smaller star systems
enriched by a finite number of SNe with full mixing of the system occurring
between each SN.

\par

Observing chemical patterns will be a challenging task. With the tools of
today and current methods for abundance analysis we are able to decrease the 
absolute observational uncertainties in the abundance ratios to $\sim 0.1$ dex. 
A strictly differential study may reach the $0.05$ dex level of uncertainty.
This may be slightly too large to allow detection of the fine-structures in 
many patterns which seem to begin appearing first at a level of about $0.05$ 
dex. However, larger uncertainties can, as we have seen, partly be compensated
for by a larger stellar sample. Also, in a number of cases patterns should be
visible already with errors in the relative abundances of $\sim 0.1$ dex.

\subsection{Existing evidence for chemical patterns}
\noindent
Our aim is here not to make a literature survey of abundance data for 
metal-poor stars or to re-analyse existing data but merely to point out some
studies and phenomena that possible can be related to variations in stellar 
yields and the formation of chemical patterns. 

\par

\begin{figure}
% \resizebox{\hsize}{!}{\includegraphics{H2954F17.eps}}
\resizebox{\hsize}{!}{\includegraphics{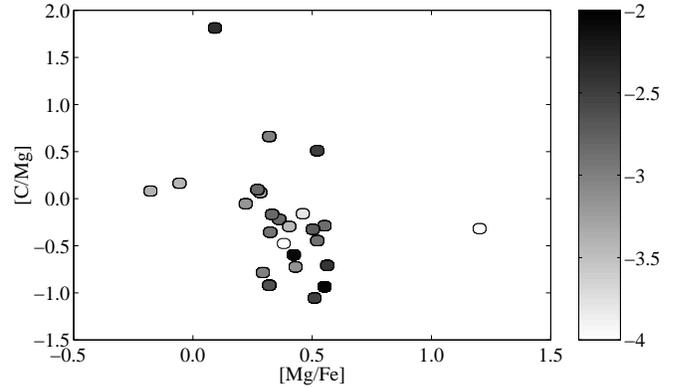}}
 \caption{Observed abundances of stars from McWilliam et al. (1995) 
  displayed in the [Mg/Fe]--[C/Mg] plane (cf. Figs. \ref{corr_diagrams}a and 
  \ref{corr_diagrams}d). The shading according to the bar on the 
  right indicates the metallicity measured by [Fe/H]. Stars with 
  $B-V>0.80$ are removed from the sample in order to 
  minimize the effect of carbon depletion in the red giants. There are too 
  few stars to allow detection of any patterns although there seems to be a 
  large, asymmetric dispersion}
 \label{MgFeCMg_obs}
\end{figure}

There may be undetected patterns in the sample of some thirty metal-poor 
giants ([Fe/H]$<-2$) observed by McWilliam et al. (1995)\nocite{mcwilliam95}. 
The sample is fairly homogeneous although the stars are not dwarfs and internal
nucleosynthesis may have altered some of their surface abundances.
Nevertheless, let us look at the distribution of stars in the [Mg/Fe]--[C/Mg] 
plane which is shown in Fig. \ref{MgFeCMg_obs}. The symbols are shaded 
according to the stellar metallicity as measured by [Fe/H]. The dispersion is 
large in both directions, spanning $1.5$ dex in [Mg/Fe] and nearly $3$ dex 
in [C/Mg]. The distribution of stars is asymmetric and there is no strong 
correlation with metallicity. However, there are too few stars in the sample 
to really allow the detection of any patterns.    

\par

\begin{figure}
% \resizebox{\hsize}{!}{\includegraphics{H2954F18.eps}}
\resizebox{\hsize}{!}{\includegraphics{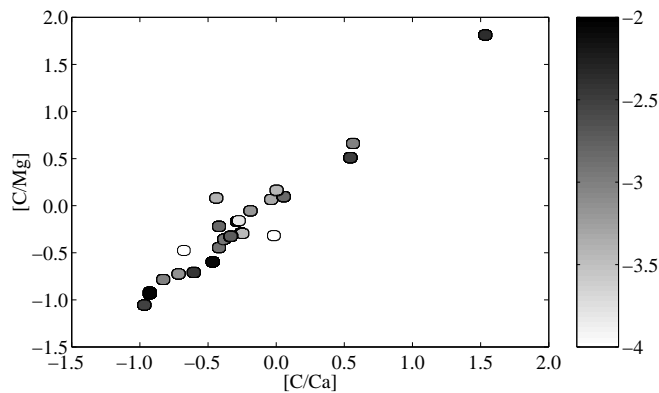}}
 \caption{Abundances of stars from the sample of McWilliam et al. (1995) 
  plotted in the [C/Ca]--[C/Mg] plane. As for Fig. \ref{MgFeCMg_obs}, the 
  symbols are shaded according to the stellar metallicity. Stars with 
  $B-V>0.80$ have been removed from the sample. Note that the 
  observed stars form a tight relation with a slope close to one}
 \label{CCaCMg_obs}
\end{figure}

If we instead plot the stars in the [C/Ca]--[C/Mg] plane, we find something 
completely different. The stars form a quite beautiful relation (see Fig. 
\ref{CCaCMg_obs}). The slope is close to unity 
and there seems to be no direct dependence on the iron abundance as indicated 
by the shading of the star symbols. In accordance with our discussion on early 
chemical enrichment we shall assume that these stars have been enriched by a
small number of core collapse supernovae at the epoch of formation of the 
Galaxy. If so, this leads us to believe that, whatever the variation of the 
yields with SN mass is, the ratio of the Ca-yield to the Mg-yield is rather
independent of progenitor mass. The scaling factor (by number relative to 
solar values) can directly be estimated from the offset in 
Fig. \ref{CCaCMg_obs} and gives $N_{\mathrm{Ca}} \simeq 1.15N_{\mathrm{Mg}}$
although it is consistent with unity. This relation can also be observed 
in the [Fe/H]--[Ca/Mg] plane where the stars are scattered around the constant 
value of $0.06$. The scaling relation between the yields (by mass) is then 
estimated to $p_{\mathrm{Ca}}(m) \simeq 0.1p_{\mathrm{Mg}}(m)$. Only weak 
observational constraints on the carbon yield can be deduced from the diagram 
in Fig. \ref{CCaCMg_obs}. However, the observed abundance ratios span almost
three orders of magnitude which is considerable.

\par

Recently, Jehin et al. (1999)\nocite{jetal99} observed a sample of mildly
metal-poor stars and found interesting correlations between the abundances of 
the $\alpha$-elements and the r- and s-process elements. The $\alpha$-elements
were correlated with the r-process elements in a one-to-one relation while
a subpopulation of the stars with high [$\alpha$/Fe] ratio seemed to be 
enriched in the s-process elements (see their Fig. 7). This excess in 
s-process elements was interpreted as an accretion phenomenon in dense 
environments, presumably globular clusters. We note that their correlation 
diagrams show strong similarities with some of the $A/A$ diagrams presented 
here, see Fig. \ref{model2_figs}a and Fig. \ref{artificial}. There is 
little doubt that the reversed "L"-shape and their two-branches-pattern are 
both caused by strong variations in the stellar yields. However, it is not 
clear whether the two-branches-pattern could be explained by yield variations 
alone or if an evolutionary effect is necessary as proposed by these authors.

\par

Another possibly related phenomenon is the existence of CN-strong and 
CN-weak line stars in some globular clusters (see e.g. Cannon et al. 
1998)\nocite{cannon98}. There seems to be a distinct bimodal abundance
pattern in this population of stars, not much different from patterns formed
by a yield with a pronounced maximum. The bimodality in the CN line strengths 
is not yet understood but, again, strong variations in the yields would 
produce similar patterns. We should emphasize that our discussion on the 
formation of patterns only holds in a strict sense for extremely metal-poor
environments while the latter two examples concern relatively metal-rich
systems.

\section{Conclusions}
\label{concl}
\noindent
Observations of the most metal-poor Galactic halo stars show convincing 
evidence for a large star-to-star scatter in abundances relative to hydrogen 
as well as abundance ratios for a variety of elements, not the least the 
neutron-capture elements (McWilliam et al. 1995\nocite{mcwilliam95}; 
Ryan et al. 1996\nocite{ryan96}; McWilliam 1998\nocite{mcwilliam98}; 
Burris et al. 2000\nocite{burris00}). This phenomenon can most easily be 
explained in terms of local enrichment of the primordial ISM by a small number 
of exploding massive stars (Audouze \& Silk 1995\nocite{as95}). Since the 
amounts of newly synthesized elements depend strongly on the mass of the 
exploding star (some elements are also affected by parameters such as rotation and metallicity), the abundances might have varied extensively throughout 
the Halo ISM. Before turbulent motions in the ISM had time to wipe out the 
chemical inhomogeneities, formation of low-mass stars occurred and the
inhomogeneities could be preserved. Hence, studying these stars 
by statistical means, especially by displaying them in $A/A$ diagrams relating
different abundance ratios, reveals important information on the production 
sites of the elements.

\par

We have demonstrated that a sample of extremely metal-poor 
stars displayed in $A/$ diagrams forms patterns which originate from specific
variations in the stellar yields. Thus, the formation of patterns is a natural
consequence of the variations in the amount of SN-produced material. 
This is clearly seen in the analytical theory. The form of the patterns 
depend critically on the shape of the integration regions which are defined
by the SN yields. Furthermore, assuming that the ejected matter from 
SN explosions can be considered chemically homogeneous (see e.g. 
Kifonidis et al. 2000\nocite{kifonidis00}; see also the comment by Arnett 
1999\nocite{arnett99}) we claim, based on the result given in 
Sect. \ref{ratio}, that chemical abundance patterns in $A/A$ diagrams 
(such as [C/Mg] vs. [Mg/Fe]) survive the effects of large-scale mixing in the 
ISM (see Fig. \ref{NaMnCo_diff}). Moreover, a comparison between the 
simulations of Model I and Model II (cf. Fig. \ref{corr_diagrams} and 
Fig. \ref{model2_figs}) indicates that the patterns are not very sensitive to 
the mass distribution function of exploding SNe (see also Fig. 3 in 
Karlsson \& Gustafsson 2000\nocite{kg00}). In the analytical expression the 
mass distribution function appears in the integrand. Hence, it merely specifies 
the relative density of stars within the patterns, not the form of the 
patterns. For a finite stellar sample, however, the apparent form of the 
patterns will depend slightly on the mass frequency of SNe since a finite 
number of stars do not cover all possible values defined by the theoretical 
density function.

\par

Continuous star formation within the clouds (Model II) could modify the 
chemical patterns. However, such an evolutionary effect would result in a 
displacement of the centre of gravity rather than the formation of new 
patterns. The number of contributing SNe in individual star-forming regions 
must be large ($>20$) in order to detect the displacement (cf. 
Figs. \ref{corr_diagrams}c, \ref{corr_diagrams}e and Fig. \ref{model2_figs}) 
and it vanishes for stars formed in a second generation of star-forming 
regions. The general characteristics of the patterns are, however, not affected 
(see Sect. \ref{ratio}).  

\par

In comparison to studies of individual Halo stars with dramatic abundance signatures like CS 22892-052, the inclusion of less extreme stars makes it possible to quantitatively discriminate between chemical patterns formed by 
different SN yields (see Fig. \ref{corr_diagrams}).
Future prospects include studies of large, homogeneous samples of extremely 
metal-poor Halo stars with accurately determined abundance ratios. In fact, 
several surveys of this kind are already under way. The idea would be to 
reconstruct the yields by solving the inverse problem. An observed chemical 
pattern in some $A/A$ plane is compared with simulations whereupon the yields 
are changed until the two distributions are, in some sense, equal. Some 
potential complications are noted. The possible contamination of thermonuclear 
SNe (SNe type Ia), intermediate-mass stars, and/or very massive stars might 
affect the chemical patterns and blur the nucleosynthetic signature of the core 
collapse SNe. Metallicity dependent yields affect the patterns of secondary 
elements. However, we estimate that for the first generations of extreme 
Pop. II stars, and for several elements, the contamination should not be very 
severe. The development of more realistic, time-dependent models should 
elucidate these problems, as well as those of mixing on various spatial and 
temporal scales. In a forthcoming paper we shall present a general method for 
SN-yield reconstruction. 

\par

A quantitative analysis is possible only if the SN yields can be 
strictly parametrized. In particular, the explosive nucleosynthesis is a 
product of a chaotic behaviour of the stellar material, intimately connected
to hydrodynamical instabilities. Shocks, convection and turbulent motions in
the SN gas could introduce intrinsic uncertainties in the produced amount of
elements. The stochastic nature of such an effect could blur the chemical 
patterns in the $A/$ diagrams. However, if the effect is sufficiently small it 
should be possible to retrieve mean yields as functions of mass, rotation etc. 
We conclude that the chemical patterns are useful diagnostics of yields of core
collapse SNe, and if patterns would be detected, we would have the opportunity
to probe the earliest phases of stellar nucleosynthesis.

\begin{acknowledgements}
We would like to thank Prof. A. Gut for several fruitful discussions on 
the theory of probability and for reading the manuscript. Valuable comments 
were also made by Dr. M. Asplund and Prof. N. Piskunov, as well as the 
anonymous referee who pointed out some crucial issues. BG acknowledge support
from the Swedish Natural Science Research Council (NFR).

\end{acknowledgements}

\begin{appendix}
\section{}
\noindent
The analytical treatment of the stellar distributions in the $A/$ diagrams
(Sect. \ref{math}) is developed in order to form a deeper understanding of the 
parameter dependence of the observed patterns. Here, we shall derive some 
general expressions for density functions as well as distribution functions 
corresponding to an arbitrary random variable or pairs of random variables.
Note, that in Sect. \ref{anal_app} we wrote the random variables in the 
subscripts within parentheses. This notation was adopted for purposes explained 
in the text. However, this in not the conventional notation, and in this 
section the subscripts are written without parentheses.

\subsection{Functions of one random variable}
\noindent
Suppose we have a random variable (r.v.) $X$ and a function $y=g(x)$. Assume
this function to be monotonically increasing with increasing $x$. Now, define 
the r.v. $Y$ as $Y=g(X)$. The probability that $Y\le y$ is then given by the
distribution function for $Y$,

\begin{eqnarray}
F_{Y}(y) & = & P(Y\le y)=P(g(X)\le y)= \nonumber\\
& = & P(X\le g^{-1}(y))=F_{X}(g^{-1}(y)),
\label{F_Y}
\end{eqnarray}

\noindent
where $F_{X}$ is the distribution function of the r.v. $X$. The inverse 
function of $g$ is denoted by $g^{-1}$. For a monotonically decreasing function
$g$ we will have that $F_Y(y)=1-F_X(g^{-1}(y))$ instead.

\par

In the continuous case the corresponding density function of $Y$ is given by
the derivative of $F_Y(y)$ with respect to $y$ such that

\begin{equation}
f_{Y}(y)=F_{X}'(g^{-1}(y))=f_{X}(g^{-1}(y)) \left| 
\frac{\mathrm{d}}{\mathrm{d}y} g^{-1}(y) \right|.
\label{generic}
\end{equation}

We can allow the function $g$ to be non-monotonic by defining it as a sum, 
$g=g_1+...+g_n$ where each function $g_i$ is monotonic and equivalent to $g$
on the open subinterval ]$x_{i-1},x_i$[ and zero elsewhere.  The 
$x_i,~i=1,...,n-1$ are real roots of $g'(x)$ and $x_0,~x_n$ are the end-points.
Thus, similarly to Eq. (\ref{generic}) the expression for $f_Y(y)$ reads 

\begin{eqnarray}
f_{Y}(y) & = & \frac{\mathrm{d}}{\mathrm{d}y}F_{Y}(y) =
\sum\limits_{i=1}^{n}\left| \frac{\mathrm{d}}{\mathrm{d}y}\left(g_i^{-1}(y)
\right) \right| f_{X}(g_i^{-1}(y)) =\nonumber\\
& = & \sum\limits_{i=1}^{n} \frac{f_{X}(g_i^{-1}(y))}{\left| 
g_i'(g_i^{-1}(y))\right|},
\label{f_Y}
\end{eqnarray}

\noindent
where $g'\equiv \mathrm{d}g/\mathrm{d}x$.

\subsection{The convolution formula}
\noindent
Given two continuous random variables, $Y$ and $Z$, with distribution 
functions $F_Y$ and $F_Z$ respectively, the sum, $X=Y+Z$, is described by the 
distribution function $F_X$ given by the integral,

\begin{equation}
F_{X}(x)=\int\!\!\!\int_{y+z \le x}f_{Y,Z}(y,z)dydz.
\label{cf_01}
\end{equation}

\noindent
If $Y$ and $Z$ are independent Eq. (\ref{cf_01}) reduces to

\begin{equation}
F_{X}(x)=\int\!\!\!\int_{y+z \le x}f_{Y}(y)f_{Z}(z)dydz.
\label{cf_02}
\end{equation}

\noindent 
This expression can be rewritten as

\begin{eqnarray}
F_{X}(x) & = & \int\limits_{-\infty}^{+\infty}f_{Y}(y)\left[
\int\limits_{-\infty}^{x-y}f_{Z}(z)dz\right]dy=\nonumber\\
& = & \int\limits_{-\infty}^{+\infty}f_{Y}(y)F_{Z}(x-y)dy.
\label{cf_03}
\end{eqnarray}

\noindent
The derivative of $F_{X}(x)$ with respect to $x$ gives the density function,

\begin{equation}
f_{X}(x) = \int\limits_{-\infty}^{+\infty}f_{Y}(y)f_{Z}(x-y)dy
\label{cf_04}
\end{equation}

\noindent
(assuming that derivation under the integral sign is allowed). 
Eq. (\ref{cf_04}) is the convolution formula for two continuous and independent
random variables.

\subsection{One- and two-dimensional distributions of $n$ random variables}
\noindent
In this section, we shall form expressions for one-dimensional density 
functions of $n$ continuous random variables as well as two-dimensional 
density functions of $n$ random variables.

\par

Given $n$ random variables $X_1,...,X_n$ and a real-valued function 
$g=g(x_1,...,x_n)$ we may form the one-dimensional r.v.

\begin{equation}
X=g(X_1,...,X_n).
\label{app_01}
\end{equation}

\noindent
For every given number $x$, we denote by $D_x$ the region in the $x_1...x_n$
hyper-plane such that $g(x_1,...,x_n) \le x$. Hence, the distribution
function of $X$ is given by the integral 

\begin{eqnarray}
F_X(x) & = & P(X \le x)=P((X_1,...,X_n) \in D_x)= \nonumber\\  
 & = &
\int...\int_{D_{x}}f_{X_1,...,X_n}(x_1,...,x_n)\mathrm{d}x_1...\mathrm{d
}x_n,
\label{app_02}
\end{eqnarray}
  
\noindent
where $f_{X_1,...,X_n}(x_1,...,x_n)$ is the density function of the joint
distribution of $X_1,...,X_n$. The corresponding density function of the r.v.
$X$ is simply given by the derivative of the distribution function $F_X(x)$,
as in Eq. (\ref{generic}).  

\par

Clearly, Eq. (\ref{app_02}) is in particular valid in the one-dimensional 
case. Namely, if that $y=g(x)$ is a monotonically increasing function, then 
Eq. (\ref{app_02}) reduces to Eq. (\ref{F_Y}).

\par

Again, suppose that we have $n$ random variables. It is then possible to 
define two new r.v.s $X$ and $Y$ via the functions $g(x_1,...,x_n)$ and 
$h(x_1,...,x_n)$ such that

\begin{equation}
 X=g(X_1,...,X_n),~~~~~~~~~~Y=h(X_1,...,X_n).  
\label{app_04}
\end{equation}

\noindent
The expression for the joint distribution of $X$ and $Y$ is similar to 
that of Eq. (\ref{app_02}). However, we shall directly express the density 
function $f_{X,Y}(x,y)$ by defining the integration region as an intersection 
between the two differential integration regions $\Delta D_x$ and $\Delta D_y$,

\begin{equation}
\{x<X\le x+\mathrm{d}x\} = \{(X_1,...,X_n) \in \Delta D_{x}\}
\label{app_05}
\end{equation}

\noindent
and

\begin{equation}
\{y<Y\le y+\mathrm{d}y\} = \{(X_1,...,X_n) \in \Delta D_{y}\}.
\end{equation}

\noindent
Note, that the density function of $X$, defined by 
Eq. (\ref{app_01}), can be calculated directly by substituting $\Delta D_{x}$ 
for $D_x$ in the integral in Eq. (\ref{app_02}). Now, for calculating the 
joint statistics of $X$ and $Y$ we introduce 
$\Delta D_{xy}=\Delta D_{x} \bigcap \Delta D_{y}$, or

\begin{eqnarray}
 \{x<X\le x+\mathrm{d}x,~y<Y\le y+\mathrm{d}y\}& = & \nonumber\\
=\{(X_1,...,X_n) \in \Delta D_{xy}\}
 \label{app_06}
\end{eqnarray}

\noindent
Finally, we then have

\begin{eqnarray}
f_{X,Y}(x,y) & = & \nonumber\\
& = & P(x < X \le x+\mathrm{d}x, y < Y \le y+\mathrm{d}y)= \nonumber\\
& = & P((X_1,...,X_n) \in \Delta D_{xy})= \nonumber\\  
 & = &
\int...\int_{D_{xy}}f_{X_1,...,X_n}(x_1,...,x_n)\mathrm{d}x_1...\mathrm{
d}x_n.
\label{app_07}
\end{eqnarray}

A general treatment of multi-variate distributions is discussed in e.g. 
Papoulis (1991)\nocite{papoulis}. Further reading on the theory of
probability can also be found in Gut (1995)\nocite{gut}.

\subsection{Some scaling laws}
\noindent
In connection to Sect. \ref{sect_convolve} and Eq. (\ref{f_nX}) we would also 
like to consider a density function describing the mean of r.v.s instead of the 
sum. With a simple scaling $f_{nX}(x)$ is transformed into

\begin{equation}
f_{\langle nX \rangle}(x) = nf_{nX}(nx),
\label{mean}
\end{equation}

\noindent
where the subscript $\langle nX \rangle$ on $f$ denotes the average of $n$ 
random variables.

\par

All functions in Sects. \ref{anal_app} and \ref{ratio} are displayed on a 
logarithmic scale. Using Eq. (\ref{F_Y}) and Eq. (\ref{f_Y}) with 
$g(X)=\log(X)$ we see that an arbitrary function of one variable transforms as 

\begin{equation}
f_{{\log X}}(x) = \ln(10) \times 10^{x}f_X(10^{x}).
\label{transform}
\end{equation}

\noindent
The factor $\ln(10)$ arises from the fact that we use the 10-logarithm, not
the natural logarithm. Similarly, a two-dimensional function transforms as,

\begin{equation}
f_{\log X,\log Y}(x,y) = \ln^2(10) \times 10^{x+y}f_{X,Y}(10^x,10^y).
\label{transform2D}
\end{equation}

\noindent
However, instead of calculating the densities on a linear scale and then
make the transformations, it is easier to directly take the logarithm of the 
generalized yields and use these functions in the calculations. 

\end{appendix}

\bibliographystyle{bibtex/apj}
\bibliography{ref}

\end{document}